\documentclass[a4paper]{article}

\usepackage[utf8]{inputenc}   
\usepackage[T1]{fontenc}      
\usepackage{geometry}         
\usepackage{graphics}
\usepackage{graphicx}
\usepackage{amsmath,amssymb}
\usepackage{color}
\usepackage{chngcntr}
\usepackage{mathrsfs} 
\usepackage{url}

\usepackage[resetlabels]{multibib}
\newcites{supp}{References}

\begin{document}

\title{Compressive three-dimensional super-resolution microscopy with speckle-saturated fluorescence excitation}   
\author{M. Pascucci$^1$, S. Ganesan$^1$, A. Tripathi$^{2,3}$, O. Katz$^2$, V. Emiliani$^1$, M. Guillon$^{1,*}$}

\newcommand{\Addresses}{{
  \bigskip
  \footnotesize

  \textsc{$^1$ Neurophotonics Laboratory UMR8250, University Paris Descartes\\
	47 rue des Saints-Pères, 75270 Paris, France}\par\nopagebreak
  \textsc{$^2$ Department of Applied Physics, The Hebrew University of Jerusalem,\\
	Jerusalem 9190401, Israel}\par\nopagebreak
	\textsc{$^3$ Department of Physics, Indian Institute of Technology,\\
	Delhi, 110016, India}\par\nopagebreak
  \textit{E-mail address} : \texttt{marc.guillon@parisdescartes.fr}

}}

\date{\today}                     

\maketitle

\Addresses

\bigskip 

\textbf{
Nonlinear structured illumination microscopy (nSIM) is an effective approach for super-resolution wide-field fluorescence microscopy with a theoretically unlimited resolution. In nSIM, carefully designed, highly-contrasted illumination patterns are combined with the saturation of an optical transition to enable sub-diffraction imaging. While the technique proved useful for two-dimensional imaging, extending it to three-dimensions (3D) is challenging due to the fading/fatigue of organic fluorophores under intense cycling conditions. Here, we present a compressed sensing approach that allows for the first time 3D sub-diffraction nSIM of cultured cells by saturating fluorescence excitation. Exploiting the natural orthogonality of transverse speckle illumination planes, 3D probing of the sample is achieved by a single two-dimensional scan. Fluorescence contrast under saturated excitation is ensured by the inherent high density of intensity minima associated with optical vortices in polarized speckle patterns. Compressed speckle microscopy is thus a simple approach that enables 3D super-resolved nSIM imaging with potentially considerably reduced acquisition time and photobleaching.
}


\bigskip 




Super-resolution fluorescence microscopy has proven to be a fundamental tool in biology to unveil processes occurring at the nano-scale~\cite{Chou_AS_11}. Sub-diffraction imaging can be obtained by exciting the fluorescent probes in a non-linear regime, either inducing binary stochastic responses~\cite{Betzig_Science_06,Zhuang_NM_06} or under saturated conditions~\cite{Hell_OL_94, Gustafsson_PNAS_05}. The saturated optical transition can be absorption~\cite{Kawata_PRL_07,Gustafsson_PNAS_05}, stimulated emission~\cite{Hell_OL_94,Hell_OE_15}, ground state depletion~\cite{Hell_PRL_07} or fluorescence photo-switching, generalized as REversible Saturable Optical Linear Fluorescence Transitions (RESOLFT) techniques~\cite{Gustafsson_PNAS_12b,Hell_NM_13}. Illumination with patterned light excitation then appears as an essential scheme~\cite{Heintzmann_JOSA_02,Betzig_NProt_14}.

In the linear excitation regime, structured illumination microscopy (SIM) with fringes~\cite{Gustafsson_JM_2000}, grids~\cite{Stallinga_OE_15} or speckles~\cite{Sentenac_NP_12, Mosk_Optica_15, Waller_BOE_17} can provide up to a two-fold improvement in resolution as compared to widefield imaging, especially in three dimensions (3D)~\cite{Gustafsson_BJ_08}. Imaging in 3D is then obtained by acquiring every single transverse plane. In this configuration, fluorescence arising from out-of-focus planes is thus induced to be wasted since suppressed by numerical sectioning~\cite{Mertz_NM_11}. Fluorescence-signal-wasting is all the more detrimental under optical-saturation conditions, required to allow breaking the diffraction limit up to theoretically unlimited resolutions, in which case non-linear photo-bleaching may occur~\cite{Gustafsson_PNAS_05,Hell_NM_07_janv}. 
Typically, the fragility of dyes has never allowed recording several transverse planes in saturated-excitation structured-illumination microscopy. 3D fluorescence nanoscopy requirements thus call for compressed sensing approaches~\cite{Candes_IEEE_08}.

Random wavefields feature two main interesting properties that make them suitable for 3D super-resolution microscopy. First, speckle patterns lying in different transverse planes are orthonormal relatively to the cross-correlation product (Section~\ref{sec:ortho}), thus allowing axial discrimination of two-dimensional objects~\cite{Marte_OE_11,Zhou_SR_18}. This property exactly provides the random-projection-measurement configuration ideally suited for compressed imaging reconstruction~\cite{Donoho_IEEE_06, Katz_APL_09}, in particular for 3D imaging~\cite{Waller_Optica_18}. 
Second, speckles exhibit strong intensity contrasts since naturally containing a high density of optical vortices of topological charge one~\cite{Berry_PRSLA_74}, such as typically used in STimulated Emission Depletion (STED) microscopy~\cite{Hell_NM_06}. Optical vortices in speckles are associated with nodal lines of intensity (in 3D) that can confine fluorescence to sub-diffraction dimensions~\cite{Pascucci_PRL_16}, despite the contribution of the axial field. 
However, whether it is possible to break the diffraction barrier with saturated speckle patterns has remained an open theoretical question~\cite{Fixler_JF_11}. 

Here, we demonstrate the possibility to achieve 3D super-resolution imaging by a single 2D raster scan under saturated fluorescence excitation with tightly focused speckle patterns. Using a custom-built speckle scanning microscope, we can image with a factor $3.3$ beyond the diffraction barrier. Super-resolution imaging capabilities are characterized using fluorescent nano-beads, and applied for imaging stained lysosomes in fixed cultured cells.  


%

\section*{Results}

\begin{figure}[!h]
\begin{center}
\includegraphics[height=0.6\textheight]{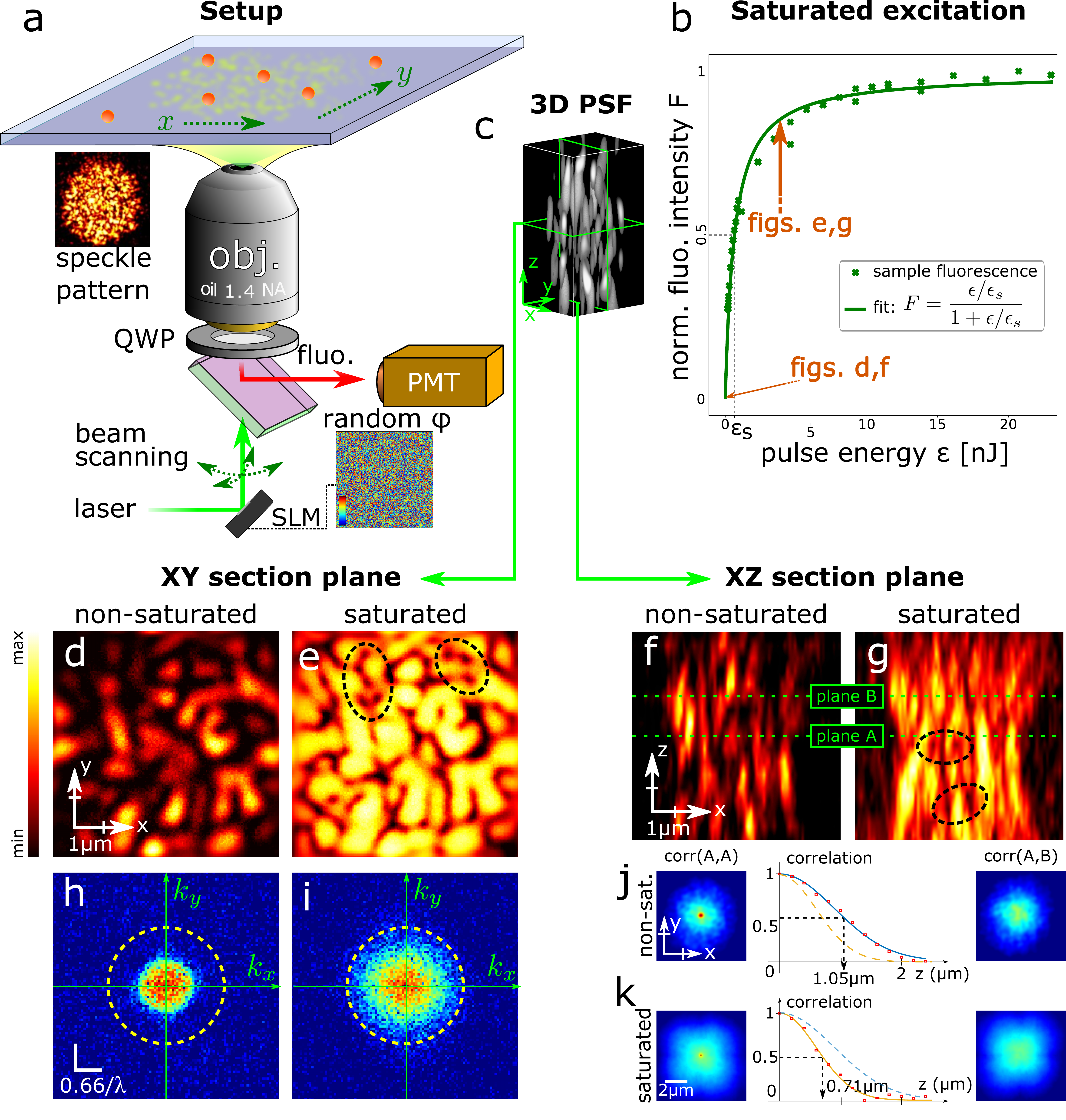}
\caption{\label{fig:setup} Principle of our speckle scanning microscope: A circularly polarized random wave-field generated by a spatial light modulator (SLM) -- displaying a random phase mask -- and a quarter wave-plate (QWP) is focused and scanned through an objective lens onto a fluorescent sample~(a). 
The fluorescence signal from a thin and dense layer of fluorescent beads illuminated with a speckle is plotted in (b), as a function of the exciting-pulse energy. 
The curve is fitted with Eq.~\eqref{eq:fit}.
The 3D speckle point spread function (SPSF) can be experimentally characterized by scanning a single fluorescent nano-bead in three dimensions (c). The speckle in the transverse plane (d,e) and in a longitudinal plane (f,g) are shown for low pulse-energy ($\left<s\right>= 5.10^{-3}$ in d,f) and for high pulse-energy ($\left<s\right>= 3.7$ in e,g) . Black dotted ellipses in d and g point out dark points identifying the plane crossing by optical vortex lines. The power spectra of SPSFs in d and e are represented in h and i, respectively, illustrating the transverse power spectrum enlargement due to saturated excitation. Cross-correlation products between different transverse planes A and B is shown in j and k, in the non-saturated and in the saturated case, respectively. In both regimes, the plots show that the cross-correlation peak vanishes with increasing defocus under saturated excitation conditions. Saturation decreases the correlation width.}
\end{center}
\end{figure}

The scanning speckle microscope is sketched in Fig.~\ref{fig:setup}a. A spatial light modulator (SLM), conjugated to the back focal plane of the objective lens, is used to generate a 3D fully developed speckle pattern (See Methods and Fig.~\ref{fig:setup_full} for a more complete description). A regular diffuser could replace the SLM but the latter allows a dynamic control of the size of the illuminating speckle pattern and thus of the intensity at the sample plane. The random wave entering the objective is circularly polarized in order to minimize the axial field at isotropic vortices of same handedness and so, to provide isotropic super-resolution in transverse planes under saturated excitation conditions (see Section~\ref{sec:isotropy}).

To quantify the saturation excitation level, we may use a two-energy-level-dye model~\cite{Enderlein_BJ_09}. The excitation probability of the dye typically depends on several parameters such as the fluorescence lifetime of the dye $\tau_f$, its absorption cross-section $\sigma$, and the laser-pulse temporal intensity profile (width $\tau_p$, amplitude $I_p$ and shape). A sub-nanosecond laser is used, delivering pluses shorter than the fluorescence life-time of the dye ($\tau_p\sim 500~{\rm ps}$) in order to efficiently saturate the optical transition with the minimal average power, and long enough for keeping a low multi-photon absorption probability. The repetition rate of $4~{\rm kHz}$ is low enough to ensure dark-state relaxation between excitation pulses and so, to minimize photo-bleaching via inter-system crossing~\cite{Hell_NM_07_janv}. When exciting fluorescence with pulses much shorter than the fluorescence lifetime ($\tau_p\ll\tau_f$), the fluorescence signal may be approximated by $F(s)\simeq 1-e^{-s}$ (See Section~\ref{sec:saturation}), where we define $s$ as the \emph{saturation parameter}. In the case of a step-wise pulse, $s=\frac{\sigma I_p\tau_p}{h\nu}$ where $h\nu$ is the quantum of excitation-light energy. For characterizing excitation saturation in our experiment, we illuminated a thin layer of beads with a speckle pattern and collected the average fluorescence signal. When averaging over intensity fluctuations of a fully developed speckle pattern (with probability density function $\rho(I)=1/\left<I\right>\exp{\left(-I/\left<I\right>\right)}$), the average fluorescence signal can be derived analytically as (Section~\ref{sec:averagefluospeckle}):
\begin{equation}
\label{eq:fit}
\left<F\right>=\frac{\left<s\right>}{\left<s\right>+1}
\end{equation} 
where the notation $\left<\cdot\right>$ stands for spatial averaging. The experimental fluorescence curve shown in Fig.~\ref{fig:setup}b is thus fitted with this function. More conveniently, the average saturation parameter $\left<s\right>$ can be expressed as $\left<s\right>=\epsilon /\epsilon_s$, with $\epsilon=I_p\tau_p A$ the pulse energy ($A$ being the speckle spot surface) and $\epsilon_s=h\nu A/\sigma$ the pulse excitation energy for which fluorescence reaches half the maximum signal. In Fig.~\ref{fig:setup}b, we measured $\epsilon_s=640~{\rm pJ}$ for a $10~{\rm \mu m}$ speckle spot.

The intensity profile of a 3D speckle pattern (Fig.~\ref{fig:setup}c) in a transverse $xy$-plane (Fig.~\ref{fig:setup}d and \ref{fig:setup}e) and in an axial $xz$-plane (Fig.~\ref{fig:setup}f and \ref{fig:setup}g) is shown both under non-saturated (Fig.~\ref{fig:setup}d and \ref{fig:setup}f) and saturated (Fig.~\ref{fig:setup}e and \ref{fig:setup}g) excitation conditions. Speckle point spread functions (SPSF) were obtained by scanning isolated fluorescent nano-beads. Under saturated illumination conditions, dark round-shaped points remain both in the $xy$ and in the $xz$ sections (highlighted by dashed ellipses in Fig.~\ref{fig:setup}e and \ref{fig:setup}g), which can be attributed to the crossing of these planes by nodal vortex lines. 
%
%
These dark points ensure contrast conservation under saturated excitation, so enlarging the power spectrum of the SPSF (Fig.~\ref{fig:setup}h and~\ref{fig:setup}i) and demonstrating the larger accessible spectral support of the optical transfer function for imaging applications. 
Contrast conservation under saturated conditions is at the basis of RESOLFT microscopy, in which resolution typically scales as~\cite{Hell_OE_08b}:
\begin{equation} 
\delta x= \frac{\lambda}{2{\rm NA}}\frac{1}{\sqrt{1+s}}
\label{eq:rescale}
\end{equation}
Although the \emph{average} saturation level in Fig.~\ref{fig:setup}d and Fig.~\ref{fig:setup}f looks modest ($\left<s\right>=3.7$) as compared with typical saturation levels used in RESOLFT microscopy, this value is averaged over intensity fluctuations of the speckle, meaning that locally, the saturation can be considerably higher. Importantly, the field gradient, 
can be large at the vortex centers where the field is minimum, ensuring efficient spatial spectrum broadening under saturated conditions. 
Breaking the diffraction limit can thus be achieved by saturating an optical transition. However, such a strategy is not always compatible with 3D imaging because of increased non-linear photo-bleaching mechanisms. In this regard, imaging with speckle is especially suited.

Indeed, 2D speckle patterns appearing in different transverse planes are statistically orthogonal relatively to the cross-correlation product. When 2D-scanning a 3D sample with a speckle, each plane of the object is convolved with a distinct pattern. The resulting image is thus the linear combination of the contribution of all the planes. Since speckles are ``quasi-orthogonal'', each planes can then be retrieved either by projection~\cite{Zhou_SR_18} or compressive sensing schemes~\cite{Waller_Optica_18}.
To be precise, speckles are only orthogonal for large-enough axial separations: $\delta z\geq 2n\lambda/{\rm NA}^2$. When saturating an optical transition, not only saturated speckles remain orthogonal but this minimal separation distance is even reduced.
The cross-correlation of two z-distant transverse planes A and B is illustrated in Fig.~\ref{fig:setup}j and \ref{fig:setup}k for the non-saturated and the saturated case, respectively, and for $z=0$ ($A=B$) and $z\gg 2n\lambda/{\rm NA}^2$. The plot of the cross-correlation product as a function of $z$ shows that the correlation distance is shorter under saturated conditions by a factor $\simeq \sqrt{2}$.  Saturating the speckle patterns further increases the axial density of modes, so-demonstrating the possibility to improve resolution along the propagation axis by optical saturation. 


\begin{figure}[!h]
\begin{center}
\includegraphics[height=0.7\textheight]{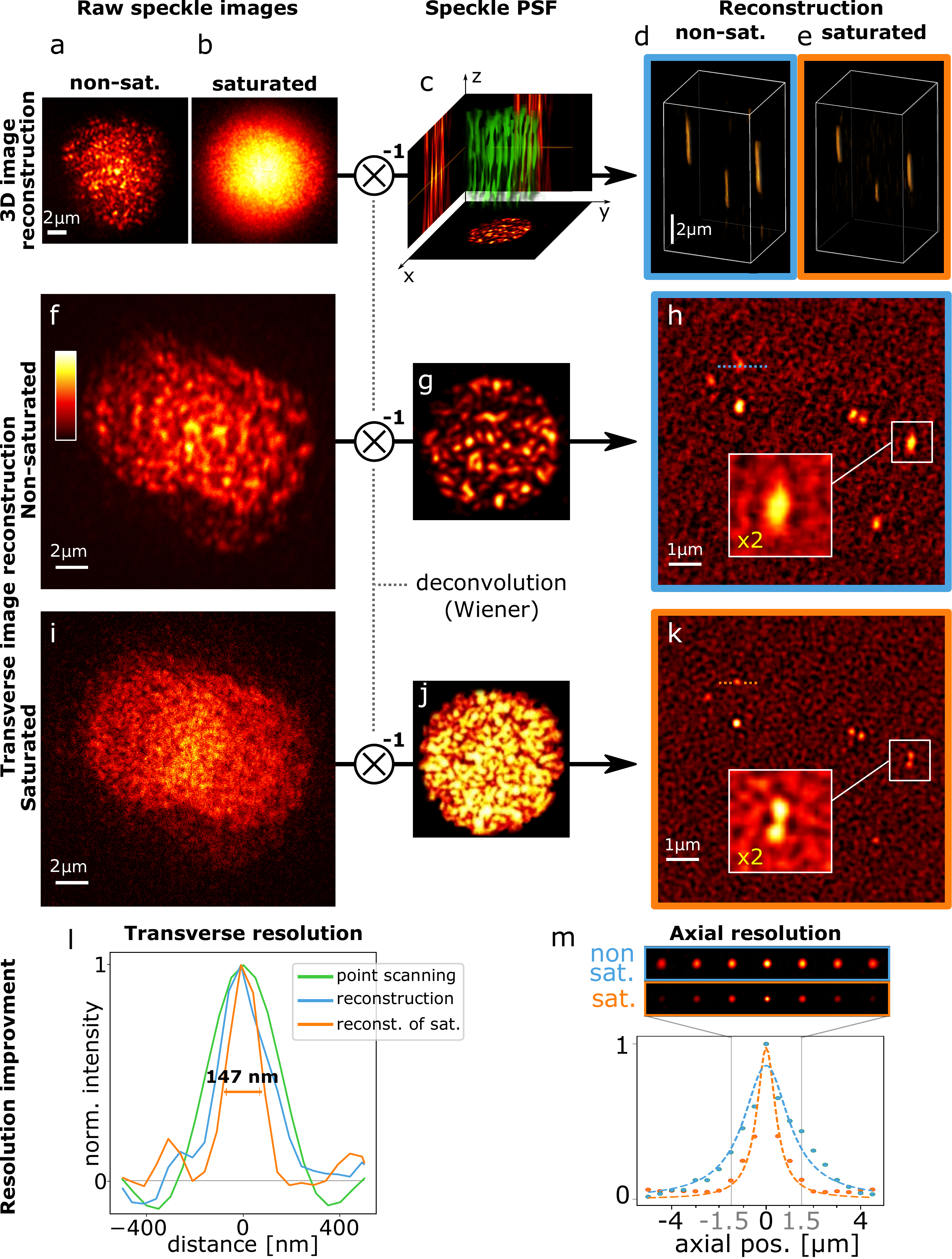}
\caption{\label{fig:deconvolution} 2D-scanning of fluorescent nano-beads by speckles (a,b) allows 3D object reconstruction (d,e). Image reconstruction is achieved by plane-by-plane Wiener deconvolution thanks to the prior experimental characterization of the 3D-SPSF (c). Speckle images were taken under non-saturated (a,f) and saturated (b,i) conditions. By depositing fluorescent $100~{\rm nm}$-beads on a coverslip, a bead cluster could be observed to be only resolved under saturated conditions (h,k). 
Line profiles in h and k are plotted in l and compared to the profile obtained by deconvolved point-scanning imaging (Fig.~\ref{fig:point_scanning}). Resolution is improved in the speckle imaging mode as compared to point-scanning mode and super-resolution is obtained under saturated excitation conditions. Axial resolution improvement is plotted in m. In all images, ${\rm NA}=0.77$ and saturated images were recorded with an average saturation parameter $\left<s\right>=1.4$.}
\end{center}
\end{figure}

Imaging with random structures raises the specific challenge of object reconstruction. Techniques have been developed to reconstruct images even when the speckles are unknown~\cite{Sentenac_NP_12,Waller_BOE_17} especially through scattering samples~\cite{Mosk_Nat_12,Psaltis_OE_14,Katz_OE_17,Min_SR_13}. Adding sparsity constraints to the rebuilt object is particularly helpful, especially to retrieve 3D information from 2D images~\cite{Waller_Optica_18}. 
Adding sparsity priors about the object in compressive sensing approaches may even provide details smaller than the resolution of the instrument~\cite{Min_SR_13}. 
Conversely, in our case, super-resolution information is experimentally extracted from the sample thanks to the power spectrum enlargement under saturated fluorescence excitation. In order to demonstrate so, 
we first perform plane-by-plane Wiener deconvolution~\cite{Fixler_OE_05,Scarcelli_SR_16}. Interestingly, Wiener deconvolution being just a cross-correlation product with a spectral renormalization, speckles satisfy the same orthogonality properties (see Section~\ref{sec:WienerX}). A single parameter must be adjusted: the mean power spectral density of the noise of images. In our results, we tune this parameter by visual image optimization. 

2D-scanned images of a sample consisting in three fluorescent beads located at different axial positions, are shown in Fig.~\ref{fig:deconvolution}a and \ref{fig:deconvolution}b, under non-saturated ($\left<s\right>\ll 1$) and saturated conditions ($\left<s\right>=1.4$), respectively. The experimental characterization of the 3D-SPSFs, required for reconstructing the object, is achieved by 3D-scanning of single isolated bead with the speckle (Fig.~\ref{fig:deconvolution}c) both in the linear and in the saturated regime. Plane-by-plane Wiener deconvolution of a 2D speckle image yields its projection on each transverse speckle plane (the 3D-SPFS) so-providing a 3D reconstruction of the object (Fig.~\ref{fig:deconvolution}d and \ref{fig:deconvolution}e). 
Since the axial correlation length is shorter in the saturated regime as shown in Fig.~\ref{fig:setup}j and ~\ref{fig:setup}k, super-resolution is also provided along the propagation axis of the beam. 
Line profiles of the beads along the axial coordinate (Fig.~\ref{fig:axial_lines}) exhibit an average axial full width at half maximum $\delta z=2.5\pm 0.1 {\rm \mu m}$ in the non-saturated regime and varies from $\delta z^\prime=1.7{\rm \mu m}$ for the bottom bead closer to the coverslip, to $\simeq 2.0 {\rm \mu m}$ (two other beads) in the saturated regime. 
Axial resolution is here better for the bead closer to the coverslip for two main reasons: - first, the index mismatch between the coverslip and the mounting medium (PVA) introduces aberrations to the speckle away from the coverslip - second, the speckle spot was focused at the coverslip, thus exhibiting higher intensities, smaller speckle grains and steeper intensity gradients near the coverslip. For this reason, for 3D imaging of biological samples presented below, we carefully focused the median plane on the median plane of the 3D sample. 
For a bead lying at the surface of the coverslip, the axial resolution is measured by deconvolving a median-plane image by all the SPSF planes (Fig.~\ref{fig:deconvolution}m). In the saturated regime the correlation distance is reduced by a factor $\simeq 2$ 
as compared to the non-saturated case. We attribute the axial resolution improvement to the presence of nodal optical vortex lines dominating the contrast (and thus the cross-correlation product) in the saturated regime, while contrast is dominated by ``speckle grains'' in the linear excitation regime. The tilted trajectory of nodal vortex lines in 3D speckle patterns~\cite{Padgett_PRL_09, Dennis_JOA_04} improves axial resolution. 


Next, a sample consisting in $100~{\rm nm}$ fluorescent beads deposited on a coverslip was prepared to measure transverse resolution improvement under saturated excitation conditions. Speckle images are shown in Fig.~\ref{fig:deconvolution}f and \ref{fig:deconvolution}i under non-saturated and saturated excitation conditions, respectively. 
The experimental measurements of SPSFs in the non-saturated and saturated cases (Fig.~\ref{fig:deconvolution}g and \ref{fig:deconvolution}j, respectively) obtained using isolated $100~{\rm nm}$ beads allow object reconstruction (Fig.~\ref{fig:deconvolution}h and \ref{fig:deconvolution}k, respectively). The image retrieved from the saturated excitation condition clearly demonstrates a higher resolution power, resolving every individual bead. The size of the bead used for the SPSF characterization limits the utmost achievable resolution, but smaller fluorescent beads yielded too low signal to background ratios. Some neighboring beads can only be resolved by saturating fluorescence excitation. Line profiles plotted in Fig.~\ref{fig:deconvolution}l show a resolution down to $\simeq 147~{\rm nm}$ in the saturated case. The width $w$ of the effective point spread function can be estimated by removing the contribution of the bead size $d$ from the width of the bead image $W$ according to $W=\sqrt{d^2+w^2}$, so giving $w=108~{\rm nm}$: a factor $3.3$ below the diffraction limit ($0.514\lambda/{\rm NA}=355~{\rm nm}$ for ${\rm NA}=0.77$). For comparison, an image was also taken by scanning a diffraction-limited spot in the sample, after correcting aberrations of the system thanks to the SLM (Fig.~\ref{fig:point_scanning}). As expected, even the resolution of the deconvolved point-scanning image is already outperformed by the image retrieved from the non-saturated speckle image by a factor $\simeq \sqrt{2}$. This improvement can be explained by the flatter optical transfer function of the speckle scanning microscope as compared to point-scanning one (see Section~\ref{sec:OTFs} and ref~\cite{Sentenac_IEEE_17}).


\begin{figure}[!h]
\begin{center}
\includegraphics[height=0.7\textheight]{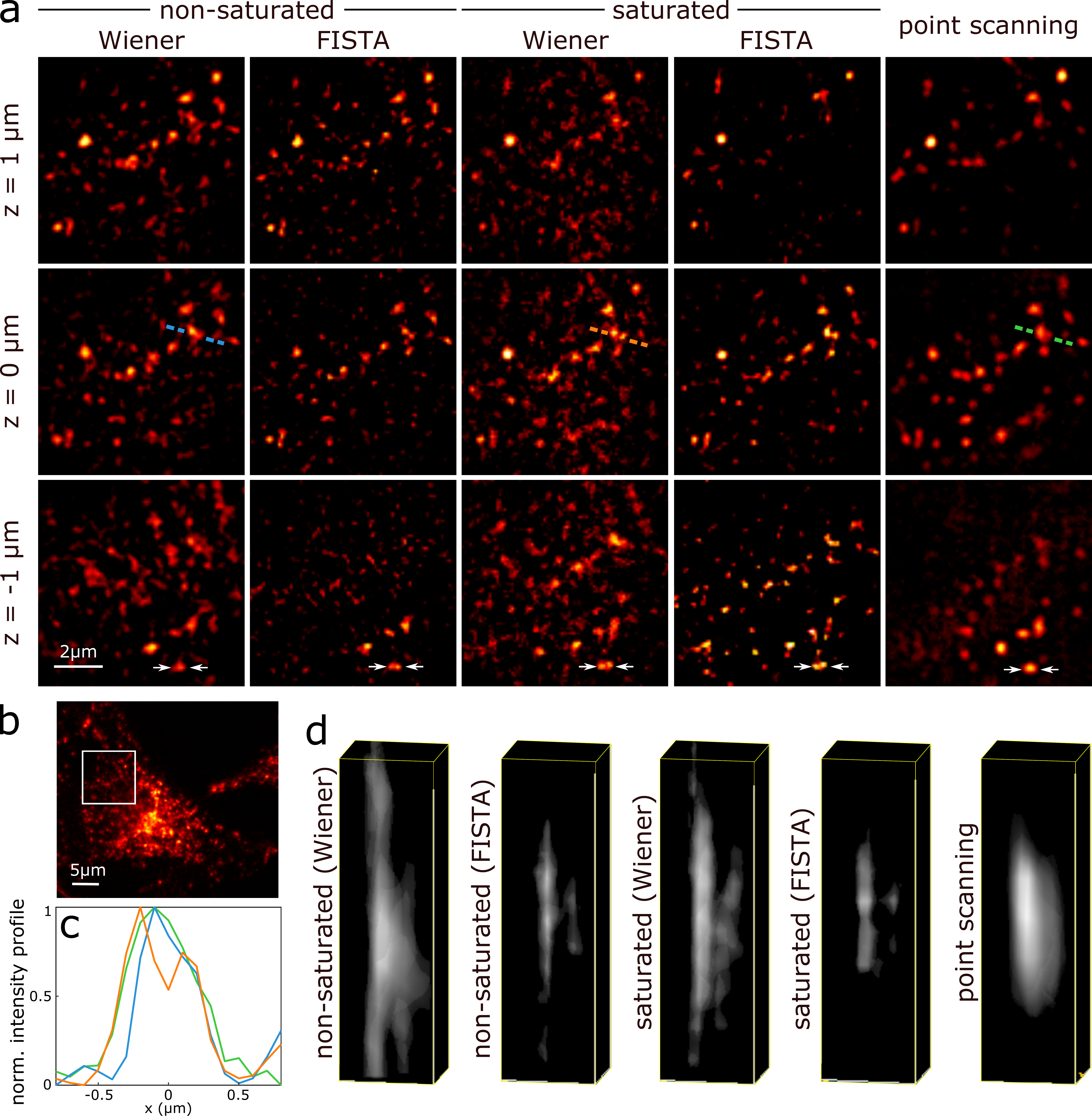}
\end{center}
\caption{\label{fig:bio} Images of lysosomes in fixed cultured cells. A region of interest materialized by a white square in the widefield view (b) was scanned both by focused spot and SPSFs. Three $1~{\rm \mu m}$-distant transverse planes from the 3D reconstruction of fluorescence distribution are shown in (a) in three imaging modes: non-saturated and saturated speckle imaging as well as 3D deconvolved point scanning. Reconstruction from speckle images is performed by Wiener deconvolution (Wiener) and by the FISTA compressed sensing algorithm. White arrows pin-point two vesicles that can only be resolved under saturated conditions. Line profiles corresponding to color dotted lines in (a) are plotted in (c). 3D intensity views of these two nearby vesicles are shown in (d). For all images ${\rm NA}=0.77$ and the average saturation parameter was $\left<s\right>\simeq 1.4$.}
\end{figure}

To demonstrate the practical applicability of 3D super-resolution speckle microscopy to biological applications, we imaged lysosomes in fixed cultured HeLa cancer cells, immuno-labeled by targeting Lysosomal Associated Membrane Protein 1 (LAMP-1). Lysosomes are vesicles whose size vary in the range between $50~{\rm nm}$ to $500~{\rm nm}$~\cite{Darchen_MBC_14}. A region of interest was first defined in a cell (Fig.~\ref{fig:bio}b). Two 2D-speckle-images (non-saturated and saturated) were recorded (not shown) and first Wiener deconvolved (Fig.~\ref{fig:bio}a, columns 1 and 3). A point scanning image-stack was also recorded and 3D-deconvolved (Fig.~\ref{fig:bio}a, last column). Three $1~{\rm \mu m}$-distant transverse planes are shown, where some nearby vesicles can only be resolved under saturated excitation conditions (see white arrows in Fig.~\ref{fig:bio}a, and line profiles shown in Fig.~\ref{fig:bio}c corresponding to dotted lines in Fig.~\ref{fig:bio}a). In speckle imaging, the depth-of-field is defined by the size of the SPSF. For Lysosomes imaging, a $5~{\rm \mu m}$ SPSF was used providing a $\sim 5~{\rm \mu m}$ depth-of-field. In Fig.~\ref{fig:actin}, a more complex object, namely actin filaments, are imaged with a $\sim 15~{\rm \mu m}$ depth-of-field using a $10~{\rm \mu m}$ speckle spot. The scaling of the depth-of-field with the speckle spot size and the numerical aperture is discussed in Sec.~\ref{sec:sparsity}.

Reconstruction by plane-by-plane Wiener deconvolution of the 2D speckle image exploits the statistical orthogonality of speckles lying in different axial planes. 
However, since speckles are only ``orthogonal'' in a statistical sense, out-of-focus point-sources contribute to a background noise whose amplitude is inversely proportional to the number of speckle grains (see Section~\ref{sec:sparsity}). 
In a worst-case scenario, a bright out-of-focus point-source may even theoretically blind a fainter one because of the out-of-focus reconstruction background. In section~\ref{sec:sparsity}, 
we demonstrate that for a scanning speckle pattern containing $N$ ``speckle grains'' in each axial plane, the sample should not exhibit more than $\simeq \sqrt{N}$ point-sources to maintain a signal to noise ratio higher than $1$ after cross-correlation projection. 
%
Moreover, even for an isolated point source, reconstruction by Wiener deconvolution yields a peak surrounded by noise. 
For these reasons, a threshold ($10\%$ of image maximum signal) was applied to Wiener-reconstructed images in Fig.~\ref{fig:bio}a to help image readability.

Such reconstruction noise artifacts can be suppressed using compressed sensing algorithms. In this regard, our speckle scanning techniques probing the sample with random point spread functions, is ideally suited~\cite{Donoho_IEEE_06,Candes_IEEE_08}. Here, a fast iterative shrinkage-thresholding algorithm (FISTA) was used~\cite{Beck_SIAM_09,FISTA_algo}. This algorithm iteratively solves such linear inverse problem as ours, by additionally taking the sample-sparsity into account. FISTA converges towards a Lagrangian minimization: 
\begin{equation}
\min_{x} \{ F(\mbox{x})\equiv ||\mbox{Ax-b}||^{2}+\lambda||\mbox{x}||_{1} \}
\end{equation}
where $\mbox{x}$ designates the object's coefficients (here, in the voxel basis where lysosomes can be described with a minimum number of coefficients), $A$ is the random projection matrix (the 3D-SPSF), $b$ the experimental 2D speckle-image and $\lambda$ a Tikhonov regularization parameter. According to compressed-sensing theory~\cite{Candes_IEEE_08}, one can reconstruct a K-sparse object (containing K non-zero coefficients) in a volume containing $N_x N_y N_z$ voxels if the number of independent random measurement points M satisfy: $M\geq O[K\log(N_x N_y N_z)]$. The results obtained by FISTA are shown in Fig.~\ref{fig:bio}a (columns 2 and 4) for comparison with Wiener deconvolution. As a result, reconstruction is achieved with a drastic reduction of noise (no threshold is applied in Fig.~\ref{fig:bio}a, contrary to Wiener reconstructions). The difference between the two reconstruction algorithms is also obvious in 3D intensity views in Fig.~\ref{fig:bio}d where the two vesicles pointed out by the dashed line in Fig.~\ref{fig:bio}a are represented. Noteworthy axial resolution is better with FISTA since less prone to yield reconstruction artifacts from out-of-focus planes, but transverse resolutions are similar in both reconstruction schemes, meaning that super-resolution is only obtained thanks to optical saturation. A more complete description as well as a quantitative characterization of sample-sparsity requirements is given in Section~\ref{sec:FISTA}, demonstrating high fidelity reconstruction by FISTA, even for quite dense objects.

\section*{Discussion}
We demonstrated that speckle patterns are ideally suited to achieve compressed 3D microscopy beyond the diffraction limit under saturated fluorescence excitation conditions. Despite the vectorial nature of light waves which disallow perfect zeros of intensity in a random light-wave structure, 
$108~{\rm nm}$ resolution could be obtained for a NA of $0.77$, a factor $3.3$ below the diffraction limit. 
Here, samples were imaged thanks to 2D raster scans with but, in principle, for sparse-enough objects, reconstruction could be achieved with even fewer measurements. This technique, based on the sparsity of the sample, opens up new perspectives to perform super-resolution imaging with potentially drastic acquisition-time reduction and photo-bleaching minimization. Contrary to regular structured illumination microscopy and confocal microscopy which wastes out-of-focus fluorescence signal, compressive 3D speckle imaging by 2D scanning makes use of all the detected fluorescence signal. 
Efficient compressive probing of 3D objects with a single 2D scanned image is ensured by the statistical orthogonality of random speckles. The cross-talk between different transverse planes can be suppressed by using a Fast Iterative Shrinkage Thresholding Algorithm (FISTA). Here, imaging of samples sparse in the voxel basis was achieved. Simple numerical deconvolution techniques could also be used for the sake of physical evidence both of 3D imaging and super-resolution capabilities. However, in practice, more complex objects could be imaged since all typical objects can be described with a sparse set of modes, provided a proper basis is used~\cite{Dahan_PNAS_12}. For objects not sparse in the voxel basis, the only limitation is the contrast of the 2D speckle image which must be larger than the photon shot-noise. Even for such objects, imaging with speckles allows efficient compressed sensing~\cite{Katz_APL_09, Delahaies_OC_11} since random structures are strongly incoherent (i.e. ``orthogonal'') with all typical sparsifying bases~\cite{Candes_IEEE_08}.


The presented results were obtained with a very simple and inexpensive system, where the microscope objective could be easily changed for a condenser of high NA, whose optical properties are poor but which can efficiently collect the fluorescence signal. In our experiments, we observed that photo-bleaching was less critical than background signal originating from the optics and the immersion oil. Here, background signal was important because fluorescence excitation was saturated but 3D super-resolution speckle imaging could also be performed saturating other optical transitions such as stimulated emission, which makes use of red-shifted light as an intense laser beam and thus both reduces background signal and photo-bleaching. In this case, speckle patterns having inverted intensity contrast~\cite{Gateau_PRL_17} could be used for the excitation and the de-excitation speckle patterns. Speckles with tailored statistical properties could also be used~\cite{Cao_OE_18}, such as non-Rayleigh speckles~\cite{Cao_PRL_14}, potentially improving statistical orthogonality. Non-diffracting speckle beams~\cite{Durnin_JOSA_87} could also allow tuning the depth of field.   

\section*{Methods}
\subsection*{The speckle-scanning microscope}
A complete scheme of the experimental setup can be found in Fig.~\ref{fig:setup_full}. For saturated speckle imaging, the laser source is a $532~{\rm nm}$ Q-switched laser diode delivering $~4\mu J$, $~500ps$ pulses at $4~{\rm kHz}$ (Teem Phononics, Meylan, France, STG-03E-120). A fully developed speckle pattern is generated by a SLM (Hamamatsu Photonics, LCOS, X10468-01) conjugated to the back focal plane of a $1.4~{\rm NA}$ microscope objective (Olympus, Tokyo, Japan, $100\times$, NA $1.40$, UPLanSApo, Oil) which focuses the beam into the sample. A quarter wave-plate is placed right before the objective lens to polarize the illuminating beam circularly. In addition, an iris placed before the SLM allows controlling the NA of the speckled beam. Finally, the speckle pattern is scanned transversely in the sample by a pair of galvanometric mirrors and fluorescence is then collected through the same objective lens and sent to a photo-multiplier tube (Hamamatsu Photonics, H10721-20). A pinhole limiting the field of view to $\sim 10\mu m$ was placed in an intermediate image plane between the objective and the photomultiplier tube to minimize background signal. The pixel dwell time corresponds to two laser pulses in 2D images and four pulses for the 3D images. In Fig.~\ref{fig:setup}f and \ref{fig:setup}g, ${\rm NA}=0.77$, and for the sake of illustration clarity, in Fig.~\ref{fig:setup}d and \ref{fig:setup}e, ${\rm NA}=0.33$. In Fig.~\ref{fig:deconvolution}f-k, $100~{\rm nm}$-beads and $50~{\rm nm}$ pixel size were used for transverse resolution characterization. For axial resolution characterization, brighter $200~{\rm nm}$ beads were used (in Fig.~\ref{fig:deconvolution}a and b) in order to help the experimentalist finding a region of interest on the camera. In Fig.~\ref{fig:bio}, 2D speckle images were recorded using a $5~{\rm \mu m}$ speckle spot with a $100{\rm nm}$ pixel size.  


\subsection*{Object reconstruction}
In Fig.~\ref{fig:deconvolution}d and \ref{fig:deconvolution}e, plane-by-plane Wiener deconvolutions were performed. The speckled point spread functions were characterized by scanning isolated $100~{\rm nm}$ fluorescent beads ($200~{\rm nm}$ for axial resolution characterization). For deconvolution of speckle images of Lysosomes in Fig.~\ref{fig:bio} (and actin filaments in Fig.~\ref{fig:actin}), both SPSFs were obtained by scanning distinct samples of fluorescent nano-beads. The saturated SPSF were estimated form these beads although the dye used was different. It appeared that the uncertainty about the exact saturation level is not critical for reconstruction. 
For Wiener deconvolution, the spectral noise density was adjusted by visual inspection in order to optimize the compromise between resolution improvement and the signal to noise ratio. In Figs.~\ref{fig:deconvolution}, deconvolved images are shown without any threshold to show the noise level. FISTA algorithm has two tuning parameters: the sparsity degree $\lambda$ and the step size $t$. For optimal reconstruction, the step size $t$ is maintained as small as possible ($10^{-7}$) and $\lambda$ is varied between $0$ and $1$; $0$ is optimal for least sparse objects and $1$ for most sparse objects. The optimization is done by comparing the root mean squared error (RMSE) \emph{a posteriori} on reconstructions using different parameters. Object reconstruction by FISTA typically took from roughly $4$ to $6$~minutes ($371.58$~seconds for $170\times 170\times 25$-voxels data and $226.54$~seconds for $150\times 150\times 25$-voxels data), performing 2000 iterations (providing good resolution as shown in Fig.~\ref{fig:bio}), on an Intel core i7-6700 processor, clocked at $3.4$~GHz.\\


\subsection*{Samples}
Fluorescent beads (FluoSpheres\textregistered, Molecular Probes, carboxylate-modified microspheres, $0.1~{\rm \mu m}$, orange fluorescent ($540/560$) $2\%$ solid) were spin-coated in a PVA matrix on a coverslip and mounted in an anti-fade mounting medium (Fluoromount, Sigma-Aldrich). 
Cultured cancer HeLa cells grown on coverslips were washed in phosphate buffer saline (PBS) and fixed with $4\%$ paraformaldehyde. Then, they were incubated in blocking solution for $30~{\rm min}$ at room temperature. Primary antibody targeting Lysosomal Associated Membrane Protein 1 (LAMP-1) were diluted in blocking buffer and incubate 1h at room temperature. After several rinces, coverslip were incubated with fluorescent Alexa 555-conjugated secondary antibody for $45~{\rm min}$ at room temperature, rinced extensively in PBS and mounted with Fluoromount anti-fading and index-matching medium.
For actin samples, monomeric G-actin was polymerized by addition of $100$~mM KCl and $2$~mM $MgCl_2$~\cite{Pernier_NSM_13}. Actin filaments were stabilized using Alexa Fluor 546 phalloidin (Thermo Fisher), and anchored on coverslips functionalized with inactive Myosin 1b~\cite{Bassereau_NC_14,Coudrier_NCB_11}. Finally, the sample was mounted in Fluoromount after several washes. 
Samples of beads in 3D were prepared by simply drying a colloidal suspension of beads with PVA. The actin filaments in 3D were mounted in Fluoromount.

\section*{Acknowledgements}
The authors thank Isabelle Fanget for preparing the HeLa cells, Julien Pernier for providing the actin filaments and the corresponding staining protocol, and Laura Caccianini, Madjouline Abou Ghali and Dany Khamsing for helping with samples. The authors also acknowledge Grace Kuo and Laura Waller for their support with the FISTA algorithm.
This work was supported by grants from the R{\'e}gion Ile-de-France, the French-Israeli Laboratory ImagiNano, and the Centre National de la Recherche Scientifique.




\newpage

\appendix
\setcounter{figure}{0}
\setcounter{section}{19}

\counterwithin{figure}{section}

\section*{Supplementary Materials}

\subsection{Experimental Setup}

\begin{figure}[!h]
\begin{center}
\includegraphics[width=\textwidth]{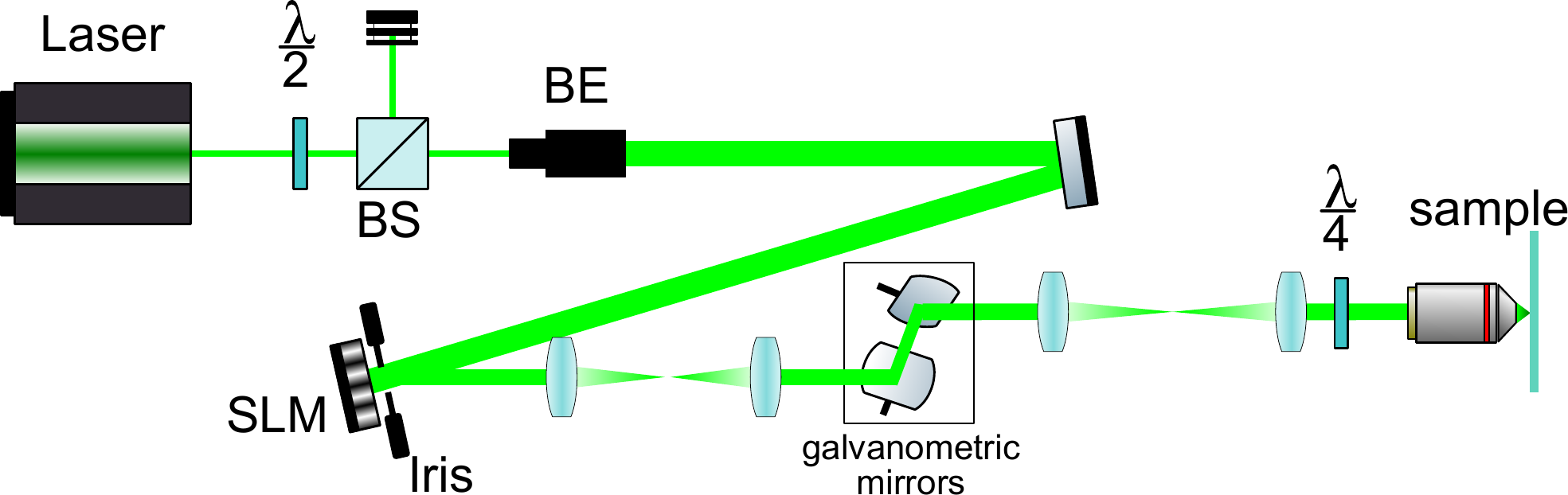}
\caption{\label{fig:setup_full} Complete scheme of the speckle scanning microscope. The laser power is modulated using a half-wave plate ($\lambda/2$) and a polarizing beam splitter (BS). The laser beam then passes through a beam-expander (BE) before illuminating the spatial light modulator (SLM) which generates the speckle. The SLM is conjugated to a pair of galvanometric mirrors and to the back focal plane of the microscope objective. A quarter wave-plate ($\lambda/4$) circularly polarizes the impinging beam in order to achieve isotropic transverse super-resolution.}
\end{center}
\end{figure}

\newpage

\subsection{Optical Transfer Function (OTF) of a scanning Microscope}
\label{sec:OTFs}

Here we derive the two-dimensional optical transfer function (OTF) of a speckle scanning microscope in the linear (i.e. non-saturated) excitation regime and demonstrate that the speckle OTF better probes the high spatial frequencies, especially in the vicinity of the OTF support boundary.

\subsubsection{Imaging system}

The two-dimensional image $I$ obtained by an optical system of a object consisting in a spatial density of fluorophore $O$ is given by a convolution with the system point spread function ${\rm PSF}$:
\begin{eqnarray}
I({\bf r}) & = & \int O({\bf r}^\prime){\rm PSF}({\bf r}-{\bf r}^\prime)d{\bf r}^\prime\\
   &=& O\ast {\rm PSF} 
\end{eqnarray}
Fourier transforming the former equation gives:
\begin{equation}
\mathscr{F} (I) =\mathscr{F} (O) \mathscr{F} ({\rm PSF})
\end{equation}
where $\mathscr{F} ({\rm PSF})$ is called the optical transfer function (OTF). The OTF thus filters the object by modulating the amplitude of its Fourier components and even canceling those lying outside the OTF support. Ideally, for the optical system to be as reliable as possible, the OTF should then be as flat as possible, at least over its support.

In the following, we compare the two-dimensional OTF of a point-scanning microscope and of a speckle-scanning microscope.

\subsubsection{OTF of a point-scanning microscope}

The PSF of a non-aberrant point-scanning microscope ${\rm PSF_0}$ is given by the (inverse) Fourier transform of the objective lens pupil $\Pi$:
\begin{equation}
{\rm PSF_0}=| \mathscr{F}^{-1} (\Pi) |^2
\end{equation}
and the corresponding OTF is then:
\begin{eqnarray}
{\rm OTF}_0 &=& \mathscr{F} \left[ \mathscr{F}^{-1} (\Pi) \mathscr{F}^{-1,\ast} (\Pi^\ast) \right]\\
 &=& \Pi \star \Pi^\ast
\end{eqnarray}
where $\star$ designate the cross-correlation product.

We now assume that the objective lens pupil is real and that it has simply a disk-shaped profile ($\Pi=1$ over this disk and $0$ outside). In the spatial-frequency domain, the radius of the disk is given by the numerical aperture (NA) of the objective lens and is equal to ${\rm NA}/\lambda$. The OTF being the cross-correlation product of this disk, the OTF profile is thus just given by the overlapping area of two disks of same radius which can be trivially analytically derived as:
\begin{equation}
{\rm OTF}_0 ({\bf w}_\perp) = 2\left(\frac{{\rm NA}}{\lambda} \right)^2\left[ \arccos\left(\frac{w_\perp\lambda}{2{\rm NA}}\right) - \frac{w_\perp\lambda}{2{\rm NA}} \sqrt{1-\left(\frac{w_\perp\lambda}{2{\rm NA}}\right)^2}   \right]
\end{equation}

\begin{figure}[!h]
\begin{center}
\includegraphics[width=10cm]{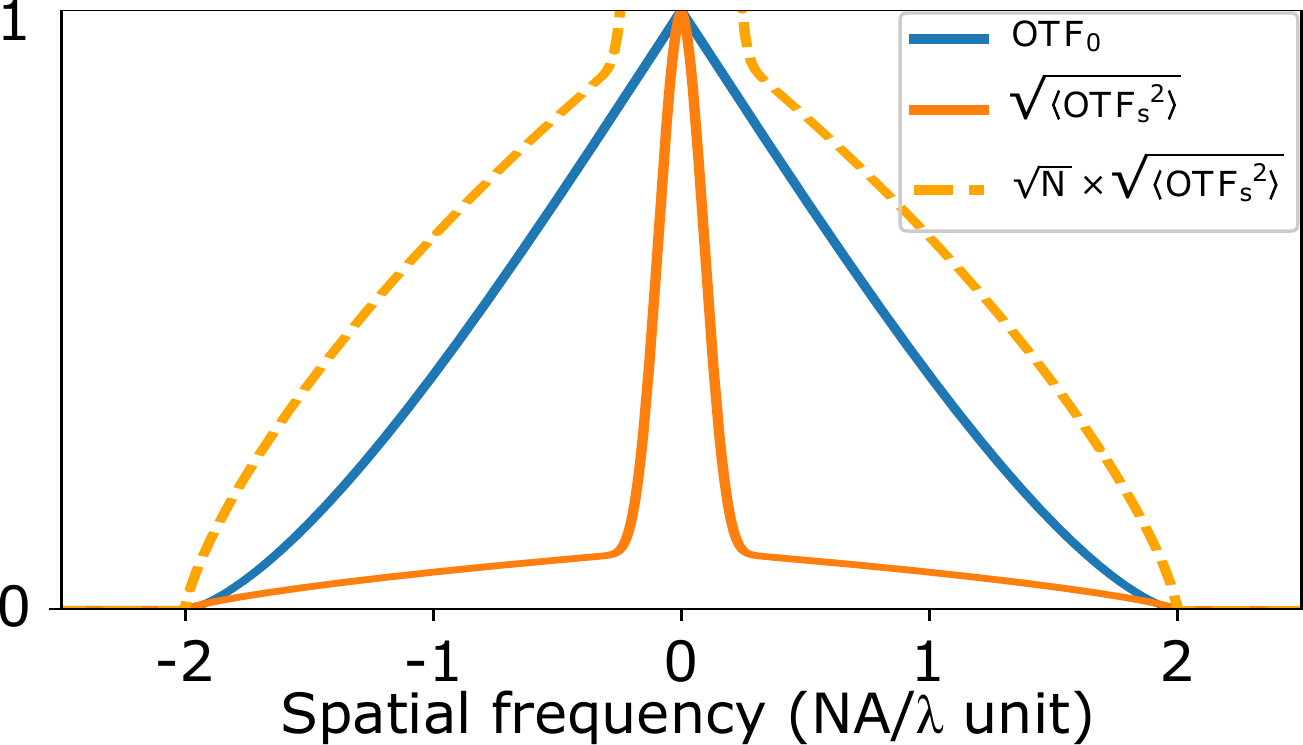}
\caption{\label{fig:OTF0} Optical transfer function of a point scanning microscope (${\rm OTF_0}$) and the mean squared OTF of a speckle scanning microscope (${\rm OTF_s}$). Above the low-frequency peak (corresponding to the inverse of the speckle spot), the speckle scanning OTF is proportional to the square root of ${\rm OTF_0}$, ensuring more efficient probing of high spatial frequencies.}
\end{center}
\end{figure}

The normalized plot of this profile is shown in Fig.~\ref{fig:OTF0}. The OTF vanishes at twice the pupil disk radius, namely $2{\rm NA}/\lambda$.

\subsubsection{OTF of a speckle-scanning microscope}

\begin{figure}[!h]
\begin{center}
\includegraphics[width=\textwidth]{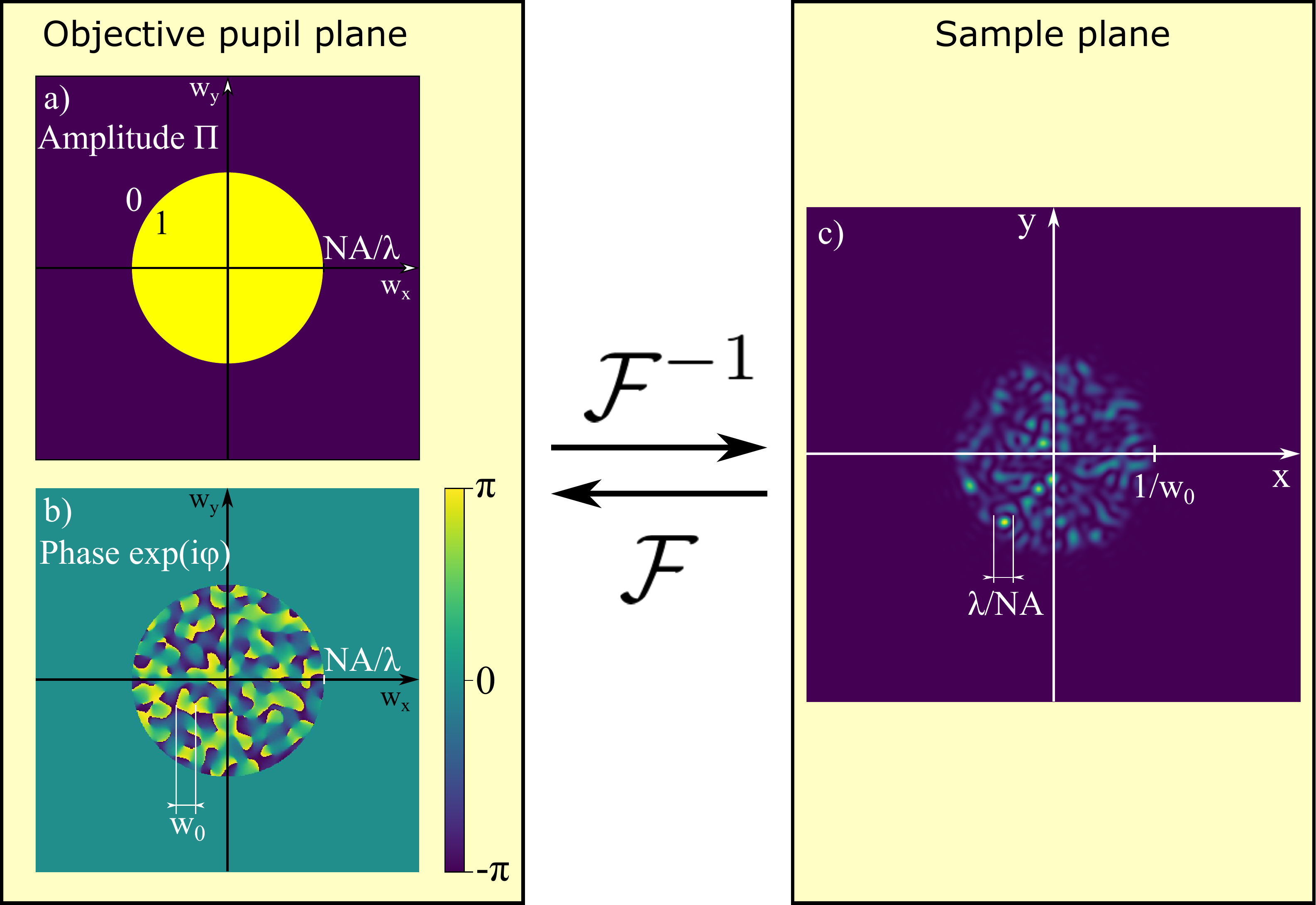}
\caption{\label{fig:OTF} Notations for the calculation of OTFs. In the back focal plane of the objective lens, the amplitude is assumed to be binary disk-shaped (a). The spatial frequency at the pupil boundary is ${\rm NA}/\lambda$. For speckle projection, a random phase is added (b). The correlation width of the field is notated $w_0$. The field at the sample plane is given by an inverse Fourier transform relation relatively to the pupil plane of the objective. The width of the resulting speckle spot (c) is $1/w_0$ and the speckle grain size is $\lambda/{\rm NA}$.}
\end{center}
\end{figure}

In our experiment, the speckle is generated by a spatial light modulator conjugated to the back focal plane of the objective lens. More generally, a speckle can be typically obtained by placing a random phase mask at the back focal plane of a lens. The field at the pupil plane may then be written $\Pi e^{i\varphi({\bf w}_\perp)}$ where $\Pi$ is the former pupil profile and $\varphi$ a random phase characterized by the covariance:
\begin{equation}
\left<e^{i\varphi({\bf w}_\perp)} e^{-i\varphi({\bf w}_\perp^\prime)}\right>=g\left(\frac{\|{\bf w}_\perp-{\bf w}_\perp^\prime\|}{w_0}\right)
\end{equation}
where $\left<\cdot\right>$ designates the statistical average, $g$ is the covariance function of the field and $w_0$ the correlation width of the random field in the pupil plane. The covariance function is peaked at value ${\bf w}={\bf 0}$ ($g(0)=1$), vanishes for large frequencies, and is characterized by a profile-width $w_0$. $w_0$ actually dictates the size of the speckle spot at the sample plane as illustrated in Figs.~\ref{fig:OTF}. The spectral extent of the pupil support being ${\rm NA}/\lambda$, we define $w_0$ so that the number of speckle grains is:
\begin{equation}
N=\left(\frac{{\rm NA}}{\lambda w_0}\right)^2
\end{equation}
Assuming $N$ is large, the statistically averaged OTF of a speckle-scanning microscope, obtained by calculating the cross-correlation product of $\Pi e^{i\varphi}$ with itself, is then trivially:
\begin{equation}
\left<{\rm OTF}_s\right> ({\bf w}_\perp) = g\left(\frac{w_\perp}{w_0}\right) {\rm OTF}_0({\bf w}_\perp)
\end{equation}
This average thus vanishes for frequencies larger than $w_0$ which means that the imaging resolution is limited by the size of the speckle spot ($1/w_0$) at the sample plane. However, prior characterization of the speckle point spread function provides the knowledge of the complex-valued OTF. Statistical complex-averaging of the OTF thus does not make sense in this case. The relevant average to calculate is then $\sqrt{\left<|OTF_s|^2\right>}$. The mean squared value of the OTF gives information about how the sample frequencies are probed by the speckle pattern. $\left<|OTF_s|^2\right>$ can be calculated the following way:
\begin{equation}
\label{eq:OTFs}
\begin{split}
&\left<|OTF_s|^2\right> \\
&= \iint \Pi({\bf w_1}) \Pi({\bf w_2}) \Pi({\bf w_1}+{\bf w}) \Pi({\bf w_2}+{\bf w}) \left< e^{-i\varphi({\bf w_1})} e^{i\varphi({\bf w_2})} e^{i\varphi({\bf w_1}+{\bf w})} e^{-i\varphi({\bf w_2}+{\bf w})} \right> d{\bf w_1}d{\bf w_2}
\end{split}
\end{equation}
According to the moment theorem for zero-mean Gaussian processes~\citesupp{Reed_IEEE_62}, the statistical average in the integral can be simplified provided that:
\begin{equation}
\left<Z_1 Z_2^\ast Z_3 Z_4^\ast\right>=\left<Z_1 Z_2^\ast\right>\left<Z_3 Z_4^\ast\right> + \left<Z_1 Z_4^\ast\right>\left<Z_3 Z_2^\ast\right>
\end{equation}
thus yielding:
\begin{eqnarray}
\left< e^{-i\varphi({\bf w_1})} e^{i\varphi({\bf w_2})} e^{i\varphi({\bf w_1}+{\bf w}_\perp)} e^{-i\varphi({\bf w_2}+{\bf w}_\perp)} \right> & = & \left< e^{-i\varphi({\bf w_1})} e^{i\varphi({\bf w_2})} \right> \left< e^{i\varphi({\bf w_1}+{\bf w}_\perp)} e^{-i\varphi({\bf w_2}+{\bf w}_\perp)} \right> \nonumber\\
& & + \left< e^{-i\varphi({\bf w_1})} e^{i\varphi({\bf w_1}+{\bf w}_\perp)} \right> \left< e^{i\varphi({\bf w_2})} e^{-i\varphi({\bf w_2}+{\bf w}_\perp)} \right> \nonumber\\
& = & g^2\left(\frac{\|{\bf w}_1-{\bf w}_2\|}{w_0}\right) + g^2\left(\frac{\|{\bf w}_\perp\|}{w_0}\right)
\end{eqnarray}
and assuming that $N\gg 1$,  Eq.~\eqref{eq:OTFs} can be simplified into:
\begin{equation}
\left<|OTF_s|^2\right> ({\bf w}_\perp) = g^2\left(\frac{\|{\bf w}_\perp\|}{w_0}\right) {\rm OTF}_0^2({\bf w}_\perp) + \left[ \int g^2\left(\frac{\|{\bf w}\|}{w_0}\right) d{\bf w} \right]{\rm OTF}_0({\bf w}_\perp)
\end{equation}
At zero frequency, where ${\rm OTF}_0$ is maximum, and assuming that $g({\bf u})=e^{-{\bf u}^2/2}$, the first term is equal to $\left[\pi\left(\frac{NA}{\lambda}\right)^2\right]^2$ while the second one is equal to $\left(\pi w_0^2\right) \left[\pi \left(\frac{NA}{\lambda}\right)^2\right] = \frac{1}{N} \left[\pi \left(\frac{NA}{\lambda}\right)^2\right]^2$ , a factor $N$ smaller. At low spatial frequencies (smaller than $w_0$), $\sqrt{\left<|OTF_s|^2\right>}$ can then be approximated by:
\begin{eqnarray}
\sqrt{\left<|OTF_s|^2\right>}  & \simeq & g\left(\frac{w_\perp}{w_0}\right) {\rm OTF}_0({\bf w}_\perp)\\
& \simeq & \left<{\rm OTF}_s\right>
\end{eqnarray}
This profile correspond to the central peak of the solid orange line in Fig.~\ref{fig:OTF0}. For frequencies larger than $w_0$, where $g$ vanishes:
\begin{eqnarray}
\sqrt{\left<|OTF_s|^2\right>}  & \simeq & \left\{ \left[ \int g^2\left(\frac{\|{\bf w}\|}{w_0}\right) d{\bf w} \right]{\rm OTF}_0({\bf w}_\perp)\right\}^{1/2}\\
&\simeq & \sqrt{\pi}w_0 \sqrt{{\rm OTF}_0({\bf w}_\perp)}
\end{eqnarray}
This profile is illustrated by the large side-lobes of orange curves in Fig.~\ref{fig:OTF0}. Since the peak amplitude of this profile is a factor $N$ as small as the central peak as explained above, the $N$-times magnified profile is shown in Fig.~\ref{fig:OTF0} as the dotted orange line. Since ${\rm OTF}_0$ vanishes linearly at its support boundary $w\leq\frac{2{\rm NA}}{\lambda}$, the speckle OTF exhibits vertical asymptotes at this boundary, ensuring more efficient probing of the object frequencies.

\subsection{Optical sectioning}
\label{sec:ortho}
\subsubsection{Orthogonality of independent speckle patterns}
Two speckle intensity appearing at different axial planes are statistically independent if the separation distance between the two planes is larger than $\simeq 2n\lambda/{\rm NA}^2$. The cross-correlation of two independent zero-mean random processes is zero. Speckled intensity are not zero-mean but the mean value can be easily removed. Based on the cross-correlation product, we may then use the following inner product for speckles~\citesupp{Freund_PhysA_90}:
\begin{equation}
\left<S_1, S_2\right>=(S_1-\left<S_1\right>)\star(S_2-\left<S_2\right>)(0)
\end{equation}
This inner product is zero for two independent speckles and for $S_1=S_2$, $\left<S_1, S_2\right>=\left< (S_1-\left<S_1\right>)^2 \right>$.
Two independent speckle are then uncorrelated according to this inner product defined from the cross-correlation product and may be said ``orthogonal''. The inner product can be calculated in the Fourier domain:
\begin{equation}
\left<S_1, S_2\right>=\mathscr{F}^{-1}\left\{ \left[\mathscr{F}(S_1)-\mathscr{F}(S_1)(0)\right]^\ast \left[\mathscr{F}(S_2)-\mathscr{F}(S_2)(0)\right] \right\}
\end{equation}

\subsubsection{Similarities beween Wiener deconvolution and cross-correlation}
\label{sec:WienerX}
Wiener deconvolution is typically performed in the Fourier domain. Assuming the OTF of the imaging system is known, Wiener-deconvolution uses the following kernel:
\begin{equation}
\label{eq:Kfunction}
K=\frac{{\rm OTF}^\ast}{|{\rm OTF}^2| +\sigma^2}
\end{equation}
where $\sigma$ prevents division by zero at locations where the OTF is smaller than the noise level or even vanishes. 
The restored object $\hat{O}$ is then obtained by:
\begin{equation}
\hat{O}=\mathscr{F}^{-1}\left[K\mathscr{F}(I)\right]
\end{equation}
Deconvolution is then very similar to a cross-correlation product. The single difference is the amplitude renormalization at the denominator of $K$ in Eq.~\eqref{eq:Kfunction}. Two speckles that are orthogonal with respect to the cross-correlation product are thus orthogonal with respect to deconvolution also.

\subsubsection{Sparsity requirements for 3D imaging by Wiener deconvolution of a 2D scan}
\label{sec:sparsity}
\begin{figure}[!h]
\begin{center}
\includegraphics[width=\textwidth]{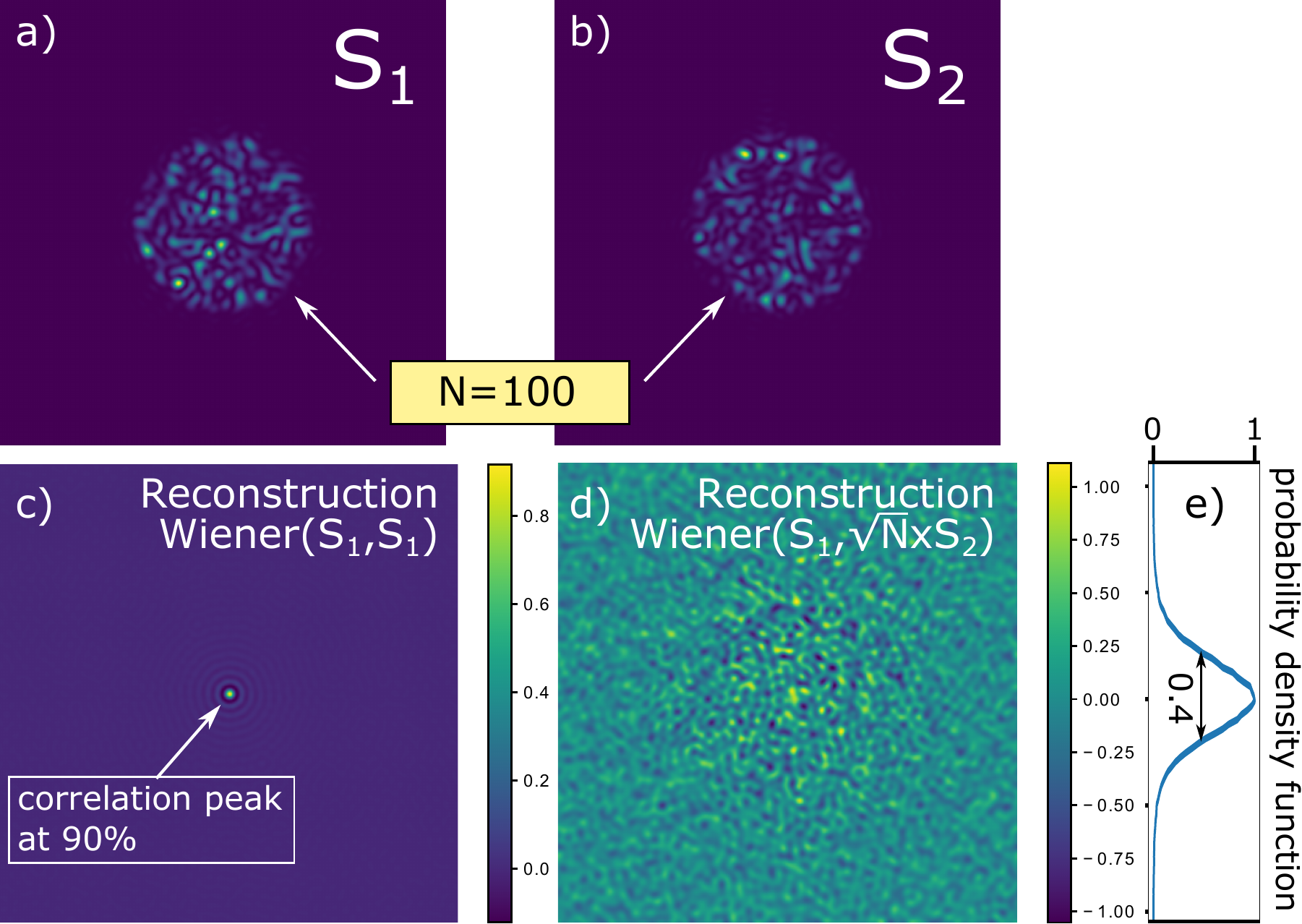}
\caption{\label{fig:sparsity} Illustration of deconvolution of two independent speckle spots (a and b) containing $N=100$ speckle grains each. Wiener deconvolution of $S_1$ by itself yields a bright centered point (c). The noise parameter in the Wiener deconvolution process was set so that the point peak reaches $0.9$. For the same noise parameter, deconvolution of $S_2$ by $S_1$ logically only results into noise whose amplitude depends on the relative average amplitude between $S_1$ and $S_2$ (d). For a $S_2$ average amplitude a factor $\sqrt{N}(=10)$ as large as the one of $S_1$, the probability density function of the noise is $0.4$ full width at half maximum (e). }
\end{center}
\end{figure}

The speckle pattern at the sample plane has a correlation length of the order of $\delta z=\frac{2n\lambda}{{\rm NA}^2}$ along the propagation axis, which means that two speckle further away than this distance are orthogonal with respect to the cross-correlation product. Moreover, the speckle grain size remains invariant over a axial range of the order of $\Delta z=2 R n/{\rm NA}$, where $R=\sqrt{N}\frac{\lambda}{NA}$ is the radius of the speckle spot. Consequently, the number of independent axial planes in this range is $N_z=\frac{\Delta z}{\delta z}=\sqrt{N}$. In the manuscript body, we deconvolve a two-dimensional scan image by every individual speckle slice of the three-dimensional speckle point spread function. Only if a point source is in this slice deconvolution yields a bright point (illustrated in Fig.~\ref{fig:sparsity}c), otherwise only noise is obtained (illustrated in Fig.~\ref{fig:sparsity}). A two-dimensional scan then seems to paradoxically provide the ability to image a volume with $N^{3/2}$ speckle grains. In the following we demonstrate that a two-dimensional scan can indeed provide a three-dimensional representation of an object, under sparsity assumptions. 

To estimate the required degree of sparsity, let's consider a point source of same brightness located at two different axial positions along the optical axis. A two-dimensional scanning of these fluorescent probes with a speckle PSF will result in the incoherent sum of two independent speckles $S_1$ and $S_2$ of same statistical properties (illustrated in Figs.~\ref{fig:sparsity}a and ~\ref{fig:sparsity}b). Deconvolution based on $S_1$ will result in a bright spot of size $\lambda/{\rm NA}$ and of amplitude close to one for the first fluorescent point source (Fig.~\ref{fig:sparsity}c). The peak amplitude depends on the noise parameter $\sigma$ chosen for deconvolution. For the second point source, deconvolution will result into noise spreading over a surface $\frac{\pi}{w_0^2}=N\pi\left(\frac{\lambda}{{\rm NA}}\right)^2$ (Fig.~\ref{fig:sparsity}d). The image reconstruction will then result in a limited signal to noise ratio defined by the signal fluctuations resulting from the deconvolution of $S_2$ by $S_1$.

To estimate this signal to noise ratio, we approximate deconvolution by a cross-correlation product. Choosing for the sake of simplicity $\left<S_1\right>=\left<S_2\right>=1$ over the speckle spot dimension, we write $S_i=1+s_i$ with $\left<s_i\right>=0$. For a fully developed speckle pattern with Gaussian statistics and Rayleigh intensity distribution, we also have $\left<s_i^2\right>=1$. Assuming that the number of speckle grains in the spot $N$ is such that $N\gg 1$, reconstruction of point source $1$ by cross-correlation with $S_1$ then yields a peak at origin with amplitude:
\begin{eqnarray}
R_{11}& =& \frac{w_0^2}{\pi}\int_{\pi/w_0^2} S_1^2 d{\bf r}\\
&=& 2
\end{eqnarray}
The cross-correlation of $S_2$ and $S_1$ yields a noisy background:
\begin{eqnarray}
R_{21}& =& \frac{w_0^2}{\pi}\int_{\pi/w_0^2} S_1 S_2 d{\bf r}\\
&=& 1+ \frac{w_0^2}{\pi}\int_{\pi/w_0^2} (s_1 + s_2) d{\bf r} +\frac{w_0^2}{\pi}\int_{\pi/w_0^2} s_1 s_2 d{\bf r}\label{eq:noisestat}
\end{eqnarray}
Statistical averaging just yields $R_{21}=1$ which correspond to a constant background. Noise must be estimated by calculating the second cumulant of $R_{21}$, resulting from the last integral in Eq.~\eqref{eq:noisestat}:
\begin{eqnarray}
\left<\left (\frac{w_0^2}{\pi}\int_{\pi/w_0^2} s_1 s_2 d{\bf r} \right)^2\right> &=& \left(\frac{w_0^2}{\pi}\right)^2 \iint_{\pi/w_0^2}\left<s_1({\bf r})s_1({\bf r}^\prime)\right>\left<s_2({\bf r})s_2({\bf r}^\prime)\right>d{\bf r}d{\bf r}^\prime\\
&=&  \left(\frac{w_0^2}{\pi}\right)^2 \iint_{\pi/w_0^2} g^2\left(\frac{\|{\bf r}-{\bf r}^\prime\|}{\lambda/{\rm NA}}\right)d{\bf r}d{\bf r}^\prime\\
&=& \left(\frac{\lambda w_0}{{\rm NA}}\right)^2\\
&=& \frac{1}{N}
\end{eqnarray}
The cross-correlation of $S_2$ and $S_1$ thus results in noise of amplitude $1/\sqrt{N}$. This noise must be compared to the $1$-amplitude peaked signal obtained by the cross-correlation of $S_1$ with itself (once the $1$-background is removed).

In the linear excitation regime, 3D object reconstruction by plane-by-plane Wiener deconvolution of a two-dimensional scan requires that the number of point sources (of same brightness) is smaller than $\sqrt{N}$ in the speckle-PSF volume with $N$ the number of speckle grains in the two-dimensional speckle PSF. Equivalently, if only two point sources of different brightness are in this volume, the dimer one can be reconstructed above the noise level if the brightness ratio between the two point sources is smaller than $\sqrt{N}$. To illustrate so, we plotted in the Wiener deconvolution result of $S_1$ by $S_1$ (Fig.~\ref{fig:sparsity}c) and of $S_1$ by $\sqrt{N} S_2$ (Fig.~\ref{fig:sparsity}c) and plotted the histogram in Fig.~\ref{fig:sparsity}e. The noise histogram is centered at zero values (since speckles are orthogonal) but exhibit fluctuations of the of $1$. Although the former analytical calculations were derived using the cross-correlation product and not Wiener deconvolution, the obtained results are qualitatively validated by numerical simulations.

Importantly, since the number of axial planes in the three-dimensional point spread function scales as $\sqrt{N}$, just like the required degree of sparsity, it must be pointed out that a line crossing the psf volume can be imaged with a signal to noise ratio equal to $1$. Uniformly fluorescent single actin filaments crossing the PSF volume can the be reconstructed by Wiener deconvolution with a signal to noise ratio of $1$.


\subsubsection{3D object reconstruction using Fast Iterative Shrinkage Thresholding Algorithm (FISTA)}
\label{sec:FISTA}

Object reconstruction by Wiener deconvolution is not optimal at least for two reasons related to the statistical orthogonality of independent speckles: - out-of-focus objects yield noise in the plane of interest and - a single point-object reconstruction is associated with noise around. In this regard, a compressed sensing algorithm is optimized to avoid these drawbacks.

The Fast Iterative Shrinkage Thresholding Algorithm (FISTA)~\citesupp{Beck_SIAM_09} is a well-known method to solve the basic linear inverse problem. It is a modification of the least square approach in which every iteration aims at minimizing the total squared error. FISTA introduces in addition a $l_1$ regularization on the object's coefficients:
\begin{equation}
\min_{x} \{ F(\mbox{x})\equiv ||\mbox{Ax-b}||^{2}+\lambda||\mbox{x}||_{1} \}
\end{equation}
In FISTA, at every iteration, matrix multiplication is followed by a ``shrinkage'' in combination with a convergence accelerator (``fast''). At every iterative step, the guessed object's coefficients are updated according to:
\begin{equation}
\mathbf{\mbox{x}}_{k+1}=T_{\lambda t}\left[\mathbf{\mbox{x}}_{k}-2t\mathbf{\mbox{A}}^{T}(\mathbf{\mbox{A}}\mathbf{\mbox{x}}_{k}-\mathbf{\mbox{b}})\right]
\end{equation}
where the shrinkage operator $T_{\lambda t}$ is given by:
\begin{equation}
T_{\lambda t}(\mbox{x})_{i}=(|\mbox{x}|-\lambda t)_{+}\mbox{sgn}(x_{i})
\end{equation}
In layman's terms, shrinkage is equivalent to what a sculptor does by iteratively chipping away small pieces from a large piece of rock. The ‘large piece of rock’ stands here for our starting guess ($x_0$) and the ‘chisel’ is the shrinkage operator. Throughout this paper we use a modified form of the FISTA implementation~\citesupp{FISTA_algo}. FISTA has two tuning parameters: the sparsity degree $\lambda$ and the step size $t$. In comparison, Wiener deconvolution has a single one, the signal to noise ratio. For optimal reconstruction, the step size $t$ is maintained as small as possible ($10^{-7}$) and $\lambda$ is varied between $0$ and $1$; $0$ is optimal for least sparse objects and $1$ for most sparse objects. The optimization is done by comparing the root mean squared error (RMSE) \emph{a posteriori} on reconstructions using different parameters. Since FISTA does not take noise into consideration, background must be subtracted and low pass filtering applied to data before running FISTA.

\begin{figure}[!h]
\begin{center}
\includegraphics[width=\textwidth]{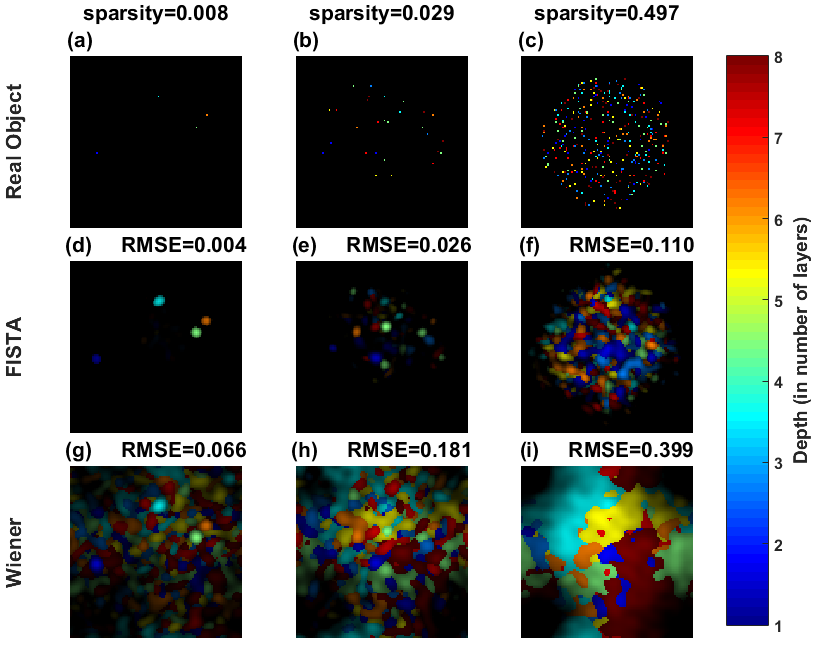}
\caption{\label{fig:FISTA1} Real objects (a)-(c), and rebuilds using FISTA (d)-(f) and Wiener deconvolution (g)-(i). Three different sparsity levels were used: $0.8\%$ ((a),(d),(g)), $2.9\%$ ((b),(e),(h)) and $49.7\%$ ((c),(f),(i)). Sparsity is defined here as the ratio between the point source density and the speckle grains density. Depth information is color-coded.}
\end{center}
\end{figure}

\begin{figure}[!h]
\begin{center}
\includegraphics[width=\textwidth]{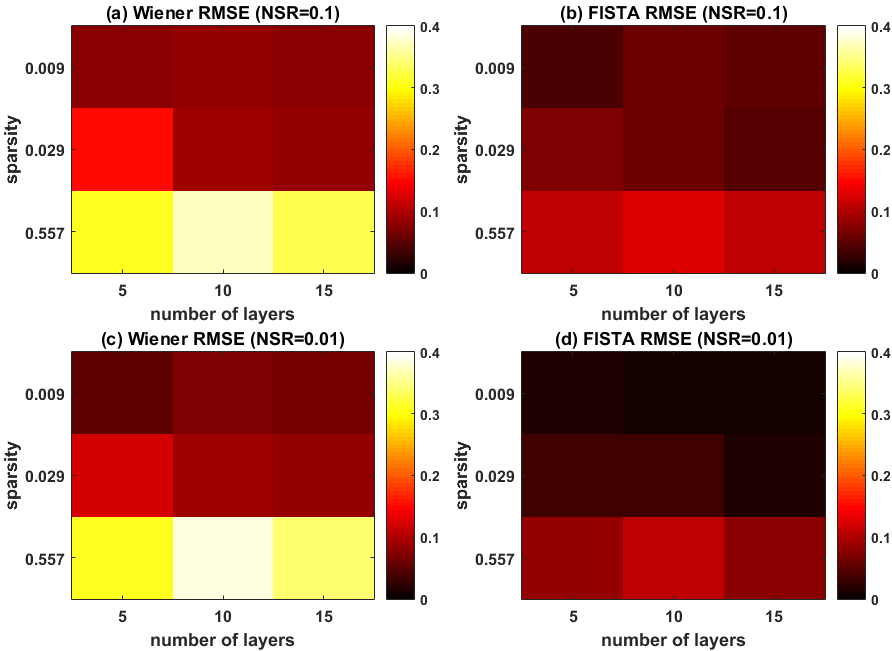}
\caption{\label{fig:FISTA2} Root mean squared error (RMSE) for reconstructions obtained by (a,c) Wiener deconvolution and (b,d) FISTA; for two different noise to signal ratios: (a,b) $NSR=0.1$ and (c,d) $NSR=0.01$. RMSE is computed for different sparsity levels and number of transverse planes.}
\end{center}
\end{figure}

In order to compare reconstructions by FISTA and Wiener deconvolution, we ran numerical simulations. In Fig.~\ref{fig:FISTA1}, three numerical objects were synthesized. Point sources of same amplitude were randomly placed in a $100\times 100\times 8$ matrix (considering $8$ transverse planes) with various sparsity degrees, so resembling the lysosome vesicles imaged in our experiment. FISTA and Wiener deconvolution were then run based on the computed 2D speckle image (not shown). Here we defined the sparsity coefficient as the ratio between the number of point sources and the number of speckle grains in the volume. With this definition, $1$-sparsity-coefficient means that the mean separation distance between point sources is $\lambda/(2NA)$ in each transverse plane (for these simulations, $\lambda/(2NA)=6.2~{\rm pixels}$). FISTA can obviously recover objects of a larger range of sparsity than the Wiener deconvolution, even though performances degrades for denser objects. Noteworthy, for dense objects, the algorithm yields better results when the initial guess is close enough to the sought-for object. In Fig.~\ref{fig:FISTA2}, the convergence of the two reconstruction algorithms were quantitatively compared by calculating the root mean squared error (RMSE) on the rebuilt 2D speckle image. Results demonstrate that FISTA clearly outperforms Wiener deconvolution and can rebuild objects up to high densities of point sources. Data reconstruction by FISTA required $96.8$~seconds to perform $2000$ iterations (providing good reconstruction) for a $100\times 100\times 10$ object on an intel i5-7500 CPU ($3.4$ GHz) with $8$ GB ram.

\newpage
\newpage
\subsection{Average fluorescence signal under short pulse excitation with speckle patterns}

\subsubsection{Fluorescence signal saturation under short pulse excitation} 
\label{sec:saturation}

Modeling of a fluorescent molecule by a two-level system~\citesupp{Enderlein_BJ_09}, the rate equation of the probability $p_1$ to be in the first excited state is:
\begin{equation}
\frac{dp_1}{dt}=k_e p_0-k_f p_1
\end{equation}
with $p_0=1-p_1$ the probability to be in the ground state, $k_e=\sigma I(t)/h\nu$ the excitation rate and $k_f=1/\tau_f$ the fluorescence rate. This differential equation can easily be solved for a step-wise excitation pulse with 
\begin{equation}
I(t)=
\left\{
\begin{array}{lcl}
I_p & {\rm for } & 0<t<\tau_p \\
0 & {\rm otherwise} &
\end{array}
\right.
\end{equation}

We then get:
\begin{equation}
p_1(t)=
\left\{
\begin{array}{lcl}
0 & {\rm for } & t<0 \\
\frac{k_e}{k_e+k_f}\left[ 1 - e^{-(k_e+k_f)t} \right] & {\rm for} &  0<t<\tau_p \\
\frac{k_e}{k_e+k_f}\left[ 1 - e^{-(k_e+k_f)\tau_p} \right] e^{-k_f t} & {\rm for} &  t>\tau_p
\end{array}
\right.
\end{equation}
Up to a multiplicative constant, the collected fluorescence signal $F$ is:
\begin{eqnarray}
F & = & \int_{-\infty}^{+\infty} k_f p_1(t) dt\\
& = & \int_{-0}^{\tau_p} k_f p_1(t) dt + \int_{\tau_p}^{+\infty} k_f p_1(t) dt\\
& = &  \frac{k_e}{k_e+k_f}\left\{ k_f \tau_p + \frac{k_e}{k_e+k_f} \left[ 1- e^{-(k_e+k_f)\tau_p} \right] \right\} \label{eq:saturation}
\end{eqnarray}
Assuming that the pulse width is much shorter than the fluorescence lifetime, the term $k_f\tau_p \ll 1$ can be neglected. Under this assumption, $F$ vanishes for low values of $k_e\tau_p$ and saturates to $1$ under intense illumination conditions, i.e. when $k_e\tau_p\rightarrow +\infty$. Under saturated speckle illumination, locations where $k_e\tau_p\ll 1$ only provide a negligible amount of signal and can be neglected. We thus only consider the case $k_e\gg k_f$. Under these two assumptions, Eq.~\eqref{eq:saturation} simplifies into:
\begin{equation}
F=1-e^{-s} \label{eq:saturation2}
\end{equation}
where $s=k_e\tau_p=\frac{\sigma I_p \tau_p}{h\nu}$ may be called the saturation parameter.

\subsubsection{Average fluorescence signal from a speckle pattern}
\label{sec:averagefluospeckle}

The excitation intensity $I_e$, and so the saturation parameter $s$, are is spatially modulated by the speckle pattern. The total fluorescence signal collected when illuminating a uniform fluorescent sample can then be calculated by averaging over the speckle intensity statistics. Assuming a fully developed speckle pattern exhibiting Gaussian statistics, the probability density function of the saturation parameter is given by:
\begin{equation}
\rho(s)=\frac{1}{\left<s\right>}e^{-s/\left<s\right>}
\end{equation}
where $\left<s\right>$ is the statistical average of $s$, which, under ergodic assumption is also the spatial average. Thanks to Eq.~\eqref{eq:saturation2}, the average fluorescence signal $\left<F\right>$ can then be calculated as:
\begin{eqnarray}
\left<F\right> & = & \int_0^\infty \rho(s)F(s) ds\\
& = & \frac{1}{\left<s\right>}e^{-s/\left<s\right>}\left(1-e^{-s}\right) ds\\
& = & \frac{\left<s\right>}{1+\left<s\right>}
\end{eqnarray}

\newpage

\subsection{Optical saturation and photobleaching}

\begin{figure}[!h]
\begin{center}
\includegraphics[width=9cm]{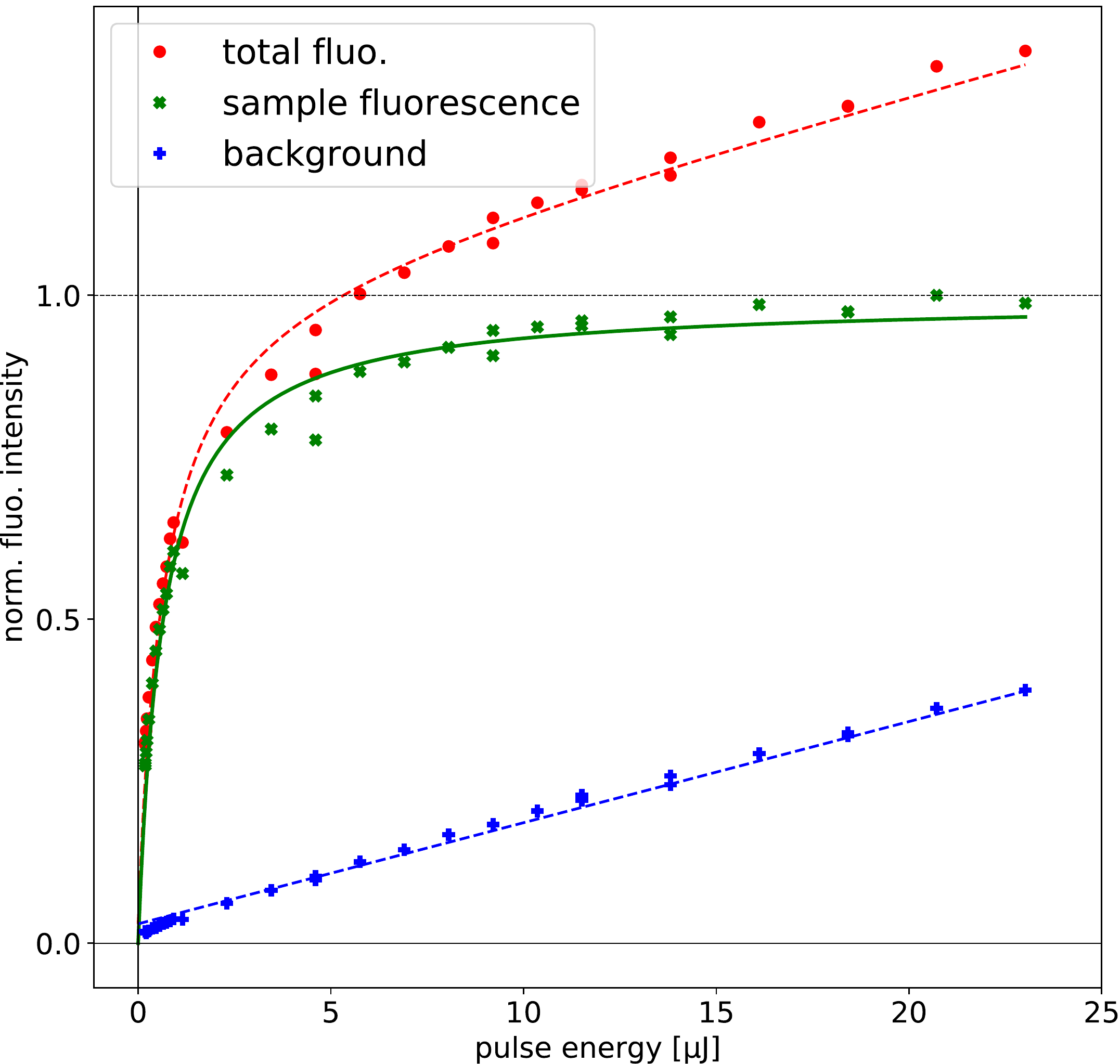}
\caption{\label{fig:saturation} Characterization of the excitation curve of the fluorescent nano-beads. The raw signal (red circles) contains both the contribution of the bead fluorescence and the background. The latter may be characterized in the absence of fluorescent bead (blue crosses). Subtracting the background to the raw signal gives the excitation curve of the fluorescent nano-bead (green x-crosses). For this experiment, a cluster of fluorescent beads was illuminated with a speckle pattern and the fitting curve thus takes into account the statistics of the intensity distribution.}
\end{center}
\end{figure}

\begin{figure}[!h]
\begin{center}
\includegraphics[width=10cm]{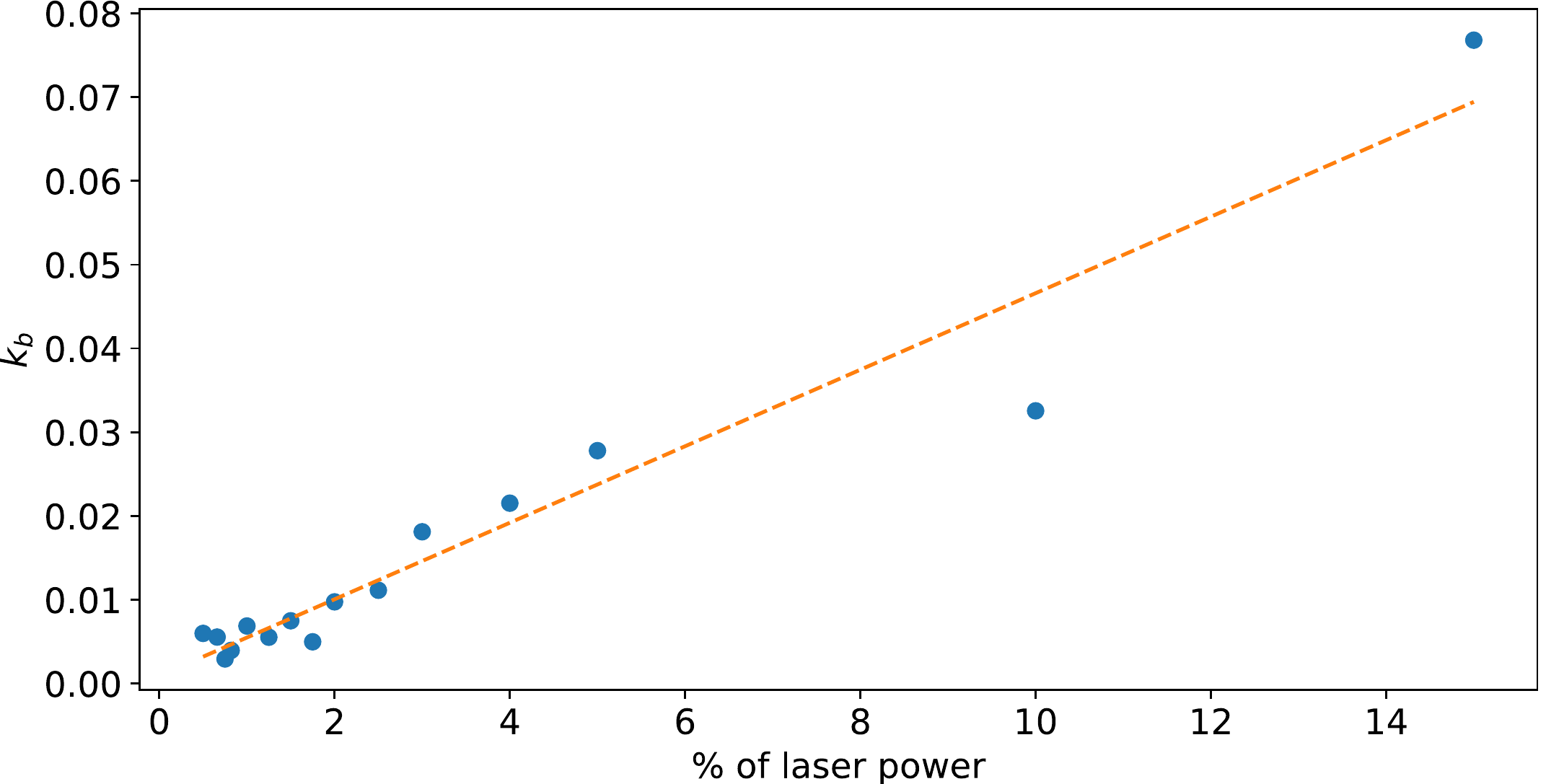}
\caption{\label{fig:bleaching} Linear evolution of photo-bleaching rate with laser intensity. Here single fluorescent $100~{\rm nm}$ nano-beads were photobleached under illumination by a focused spot of $0.22~{\rm NA}$.}
\end{center}
\end{figure}

\newpage

\subsection{Axial field modulation at the center of vortices by polarization control}
\label{sec:isotropy}

\begin{figure}[!h]
\begin{center}
\includegraphics[width=\textwidth]{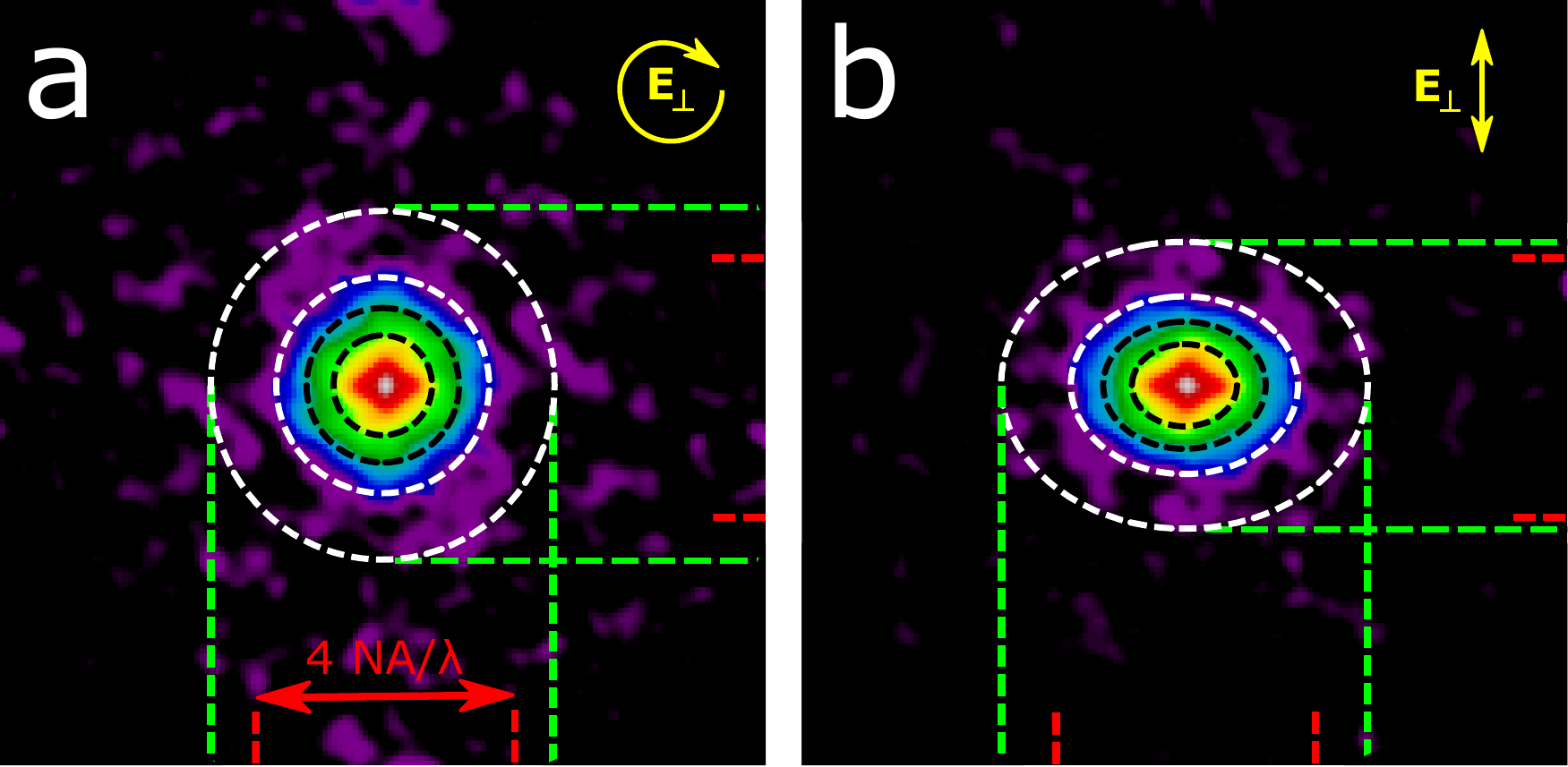}
\caption{\label{fig:anisotropic} Effect of the polarization state of the speckle pattern on the power spectrum enlargement of the speckle scanning fluorescent image. Circular polarization (a) provides isotropic power-spectrum enlargement while a vertically polarized speckle pattern minimizes the axial field at vortices strongly elongated along the vertical direction, thus enlarging the power spectrum along the horizontal direction (b). Power spectra obtained using ${\rm NA}=0.77$ and an average saturation parameter $\left<s\right>=1.4$.}
\end{center}
\end{figure}

Light being a vector wave, in existing implementation of super-resolution techniques based on the saturation of an optical transition, the pattern is prepared in a specific polarization state~\citesupp{Liu_JO_10,Bianchini_OE_12,Lounis_OE_14} in order to yield simultaneous cancellation of all three vector components of the field and to maximize contrast. In purely random waves, the optical vortices of the three vector components are unlikely to overlap, thus preventing the existence of intensity zeros.
The axial field amplitude at the center of vortices (of the transverse components) depends on their topological characteristics and on the polarization state of the beam. 
More in details, vortices in random waves are primarily characterized by their topological charge ($\pm 1$), and at first order, can be described by their elliptical intensity profile around the phase singularity~\citesupp{Berry_PRSLA_2000}. This elliptical profile in intensity is associated with an elliptically non-uniform increase of the phase along the azimuthal coordinate around the phase singularity. 
Vortices are then described by six parameters~\citesupp{Freund_JOSA_94} whose geometrical ones are the eccentricity of the ellipse and its orientation. The analytical expression and the numerical study of probability density functions of these parameters in random waves have been thoroughly discussed in the literature~\citesupp{Freund_JOSA_94,Shamir_JOSA_96, Freund_JOSA_97, Berry_PRSLA_2000}. 
To perfectly cancel the axial field at the center of a vortex (of the transverse component), polarization with the same eccentricity and the same axes as the intensity ellipse must be chosen. In a polarized random wavefield, circular polarization thus optimizes the darkness of isotropic vortices of same handedness~\citesupp{Pascucci_PRL_16} (exhibiting a uniform phase increase along the azimuthal coordinate) 
so-ensuring isotropic power-spectrum broadening by optical saturation. An illustration of the anisotropic power spectrum broadening when using linearly polarized light is shown in Fig.~\ref{fig:anisotropic}. 

\begin{figure}[!h]
\begin{center}
\includegraphics[width=\textwidth]{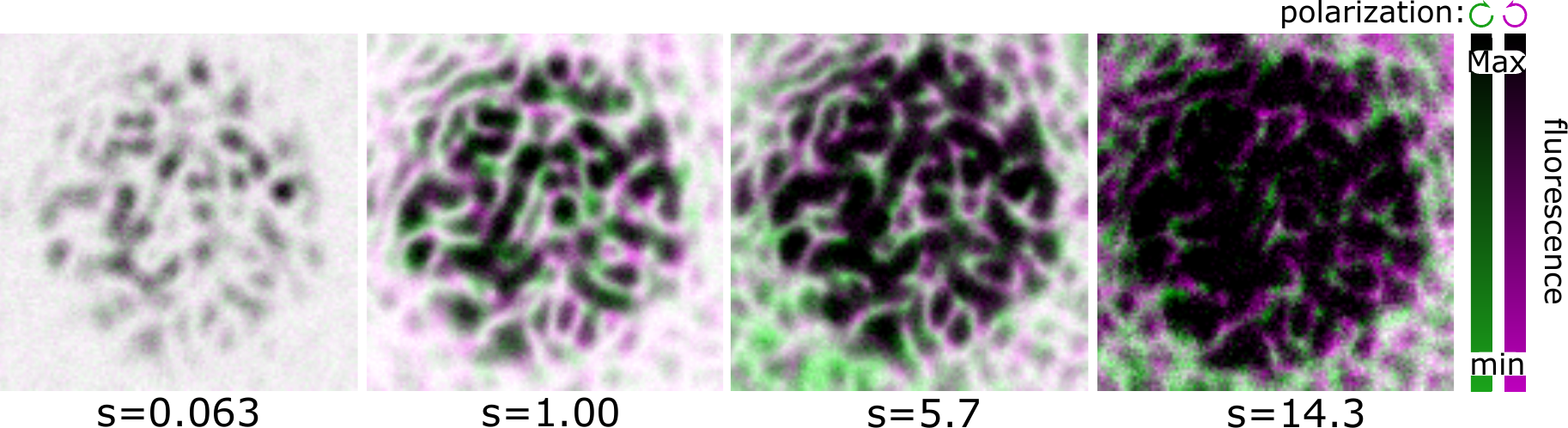}
\caption{\label{fig:polar} Effect of the axial field on the saturated fluorescence signal. In each image, the green and magenta images are obtained using the same random phase mask on the SLM for excitation, but having right and left-handed circular polarizations, respectively. Changing the handedness of circular polarization mostly modulates the axial field. Here, the contrast of images have been inverted as compared to usual representation of intensities, in order to better visualize the contribution of the axial field. Bright pixels thus code for the dark regions of the speckle which are crucial for super-resolution imaging. From left to right, the saturation parameter $s$ is increased. The significant difference observed between the green and the magenta image observed at large saturation parameters demonstrates the high sensitivity to the axial field. Images taken using a $5~{\rm \mu m}$ speckle spot, with ${\rm NA}=0.77$.}
\end{center}
\end{figure}

In a polarized random wavefield focused with a lens, the vortices of the transverse components coincide. Here, we discuss how the axial field at the center of these vortices depends on the polarization. Without loss of generality, let us choose Cartesian coordinates centered on a given optical vortex and aligned with the main axes of its characteristic ellipse. At first order development, the transverse field may then be written: 
\begin{equation}
{\bf E_\perp}=\left(\frac{x}{a}+i\sigma\frac{y}{b}\right) \left(\cos\theta {\bf e_x}+e^{i\varphi}\sin\theta {\bf e_y}\right)
\end{equation}
where $a,b>0$ are the semi-minor and semi-major axes of the ellipse, $\sigma=\pm 1$ is the topological charge of the vortex, $\theta$ the angle of the polarization ellipse with respect to the $x$-axis and $\varphi$ the relative phase between the $x$ and $y$ components of the transverse field. Using the Maxwell-Gauss equation ($\nabla {\bf E}={\bf 0}$), we obtain the axial field in the paraxial approximation:
\begin{equation}
E_z=\frac{1}{ik}\left(\frac{\cos\theta}{a}+i\sigma e^{i\varphi}\frac{\sin\theta}{b}\right)
\end{equation}
where $k$ is the wavenumber.
The axial field is thus canceled at the vortex center of ${\bf E_\perp}$ if the beam is elliptically polarized with the same handedness ($e^{i\varphi}=i\sigma$), the same ellipticity ($\tan\theta=b/a$) and the same orientation as the vortex. Since vortices in random waves contain a broad statistical distribution of ellipticities, intensity cannot be canceled at all vortices at once. The critical role of the polarization state, and thus of the axial field, at high NA and high saturation levels was demonstrated in~\citesupp{Pascucci_PRL_16} and is illustrated in Fig.~\ref{fig:polar}. Typically, a linearly $x$-polarized beam minimizes intensity at vortices strongly elongated along the $x$ dimension, and circular polarization minimizes intensity at vortices of same handedness~\citesupp{Pascucci_PRL_16}. An illustration of anisotropic power spectrum broadening is shown in Fig.~\ref{fig:anisotropic}. For imaging application, optimization of isotropic vortices is preferable in order to obtain isotropic super-resolution in the transverse plane.\\

\newpage

\subsection{Resolution improvement by optical saturation}
\begin{figure}[!h]
\begin{center}
\includegraphics[width=\textwidth]{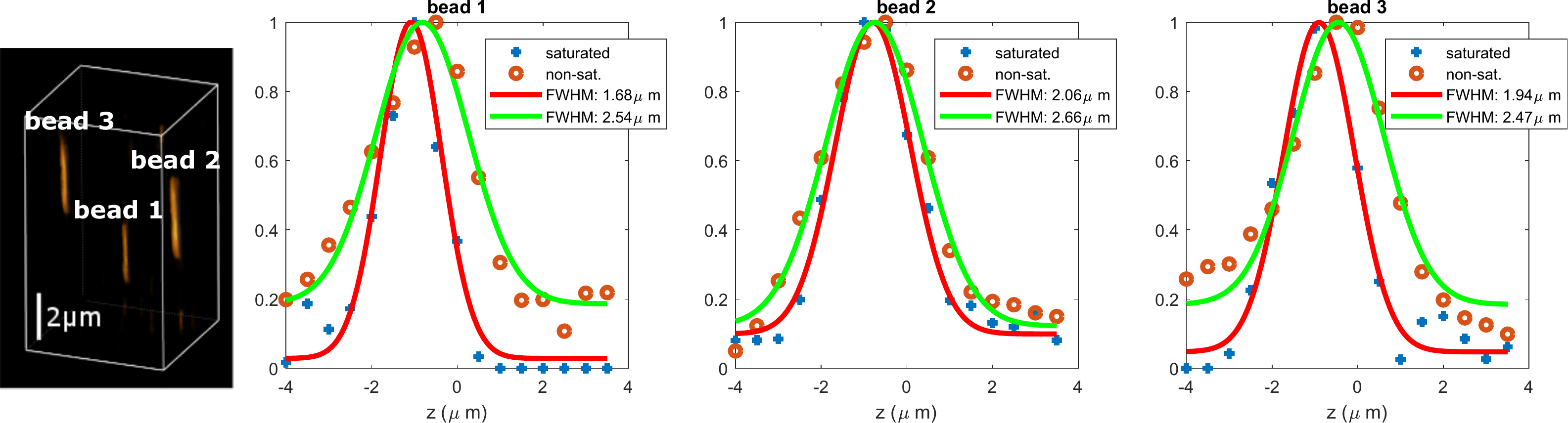}
\caption{\label{fig:axial_lines} Axial resolution improvement. Axial line profiles of beads labeled from $1$ to $3$ in the non-saturated regime (red circles) and in the saturated (blue cross) regimes. Full widths at half maxima are mesured by Gaussian fitting (solid lines).}
\end{center}

\end{figure}
\begin{figure}[!h]
\begin{center}
\includegraphics[width=12cm]{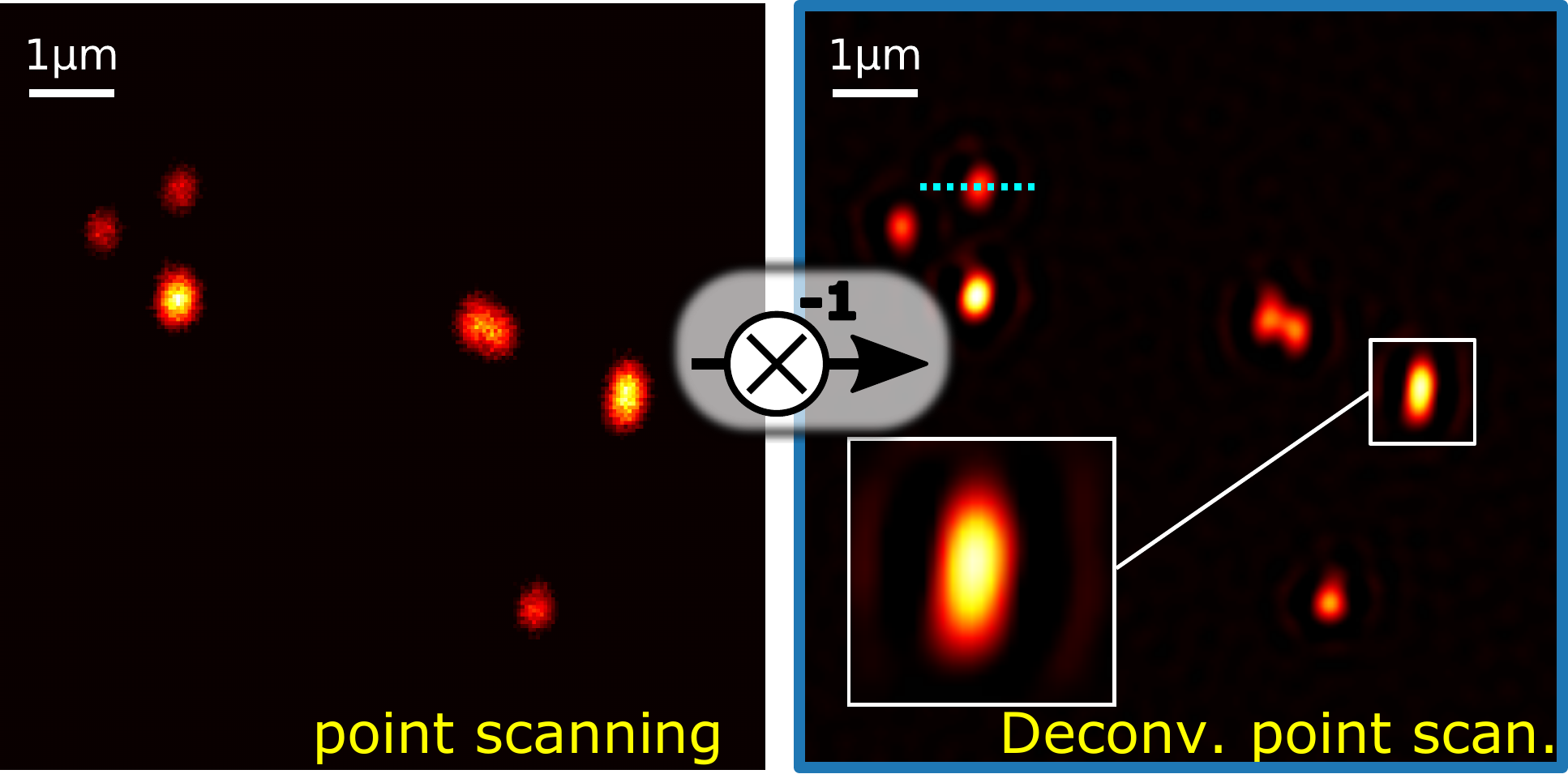}
\caption{\label{fig:point_scanning} Point scanning image (left) of the sample shown in Fig.~\ref{fig:deconvolution}f,h,i and k and its corresponding deconvolution (right) by a Gaussian fit of the experimental point spread function.}
\end{center}
\end{figure}

\newpage

\subsection{Image of actin filaments}

\begin{figure}[!h]
\begin{center}
\includegraphics[width=\textwidth]{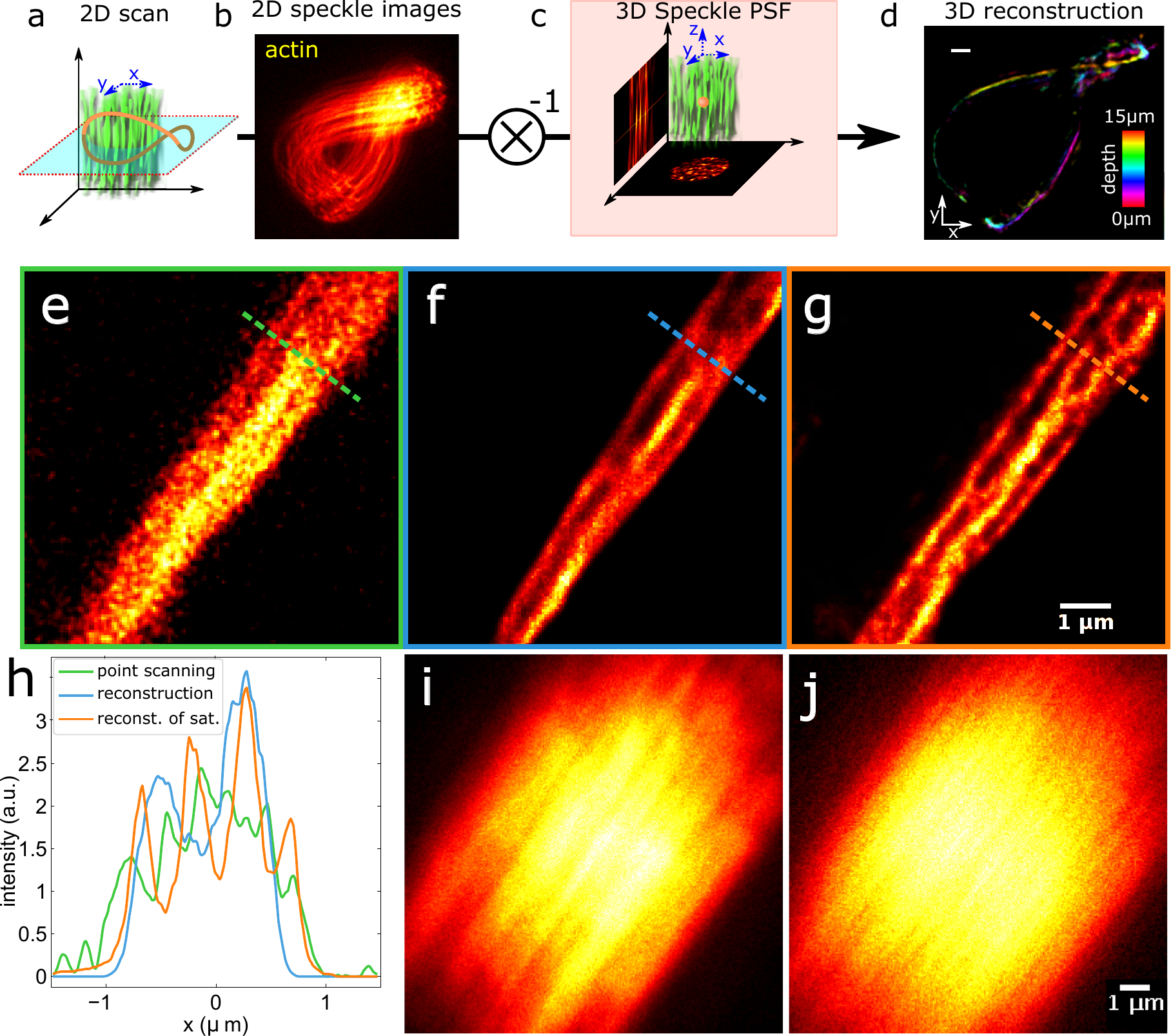}
\caption[]{\label{fig:actin} The 2D speckle images of actin filaments are shown in (b). Prior recording of the 3D-SPSF with an isolated fluorescent beads (c) and (Wiener) deconvolution of 2D-scans allows 3D reconstruction of the objects (d). In (d), the axial depth of the 3D-image of the actin filament is color-coded over a $15~{\rm \mu m}$ axial range. Scale bars in (d): $2~{\rm \mu m}$. Point-scanning image of actin filaments attached on a coverslip (e) and images reconstructed from linear (f) and saturated (g) speckle images. 1000 iterations of an iterative Richardson-Lucy algorithm~\citesupp{Unser_Meth_17} were run to reconstruct speckle images. Line profiles corresponding to the dotted lines in (e), (f) and (g) are plotted in (h). Raw speckle images corresponding to images (f) and (g) are shown in figures (i) and (j), respectively. All speckle images were recorded using a $10~{\rm \mu m}$ speckle spot with ${\rm NA}=0.77$.}
\end{center}
\end{figure}


\newpage

\subsection{Wiener and Richardson-Lucy deconvolution} 

\begin{figure}[!h]
\begin{center}
\includegraphics[width=12cm]{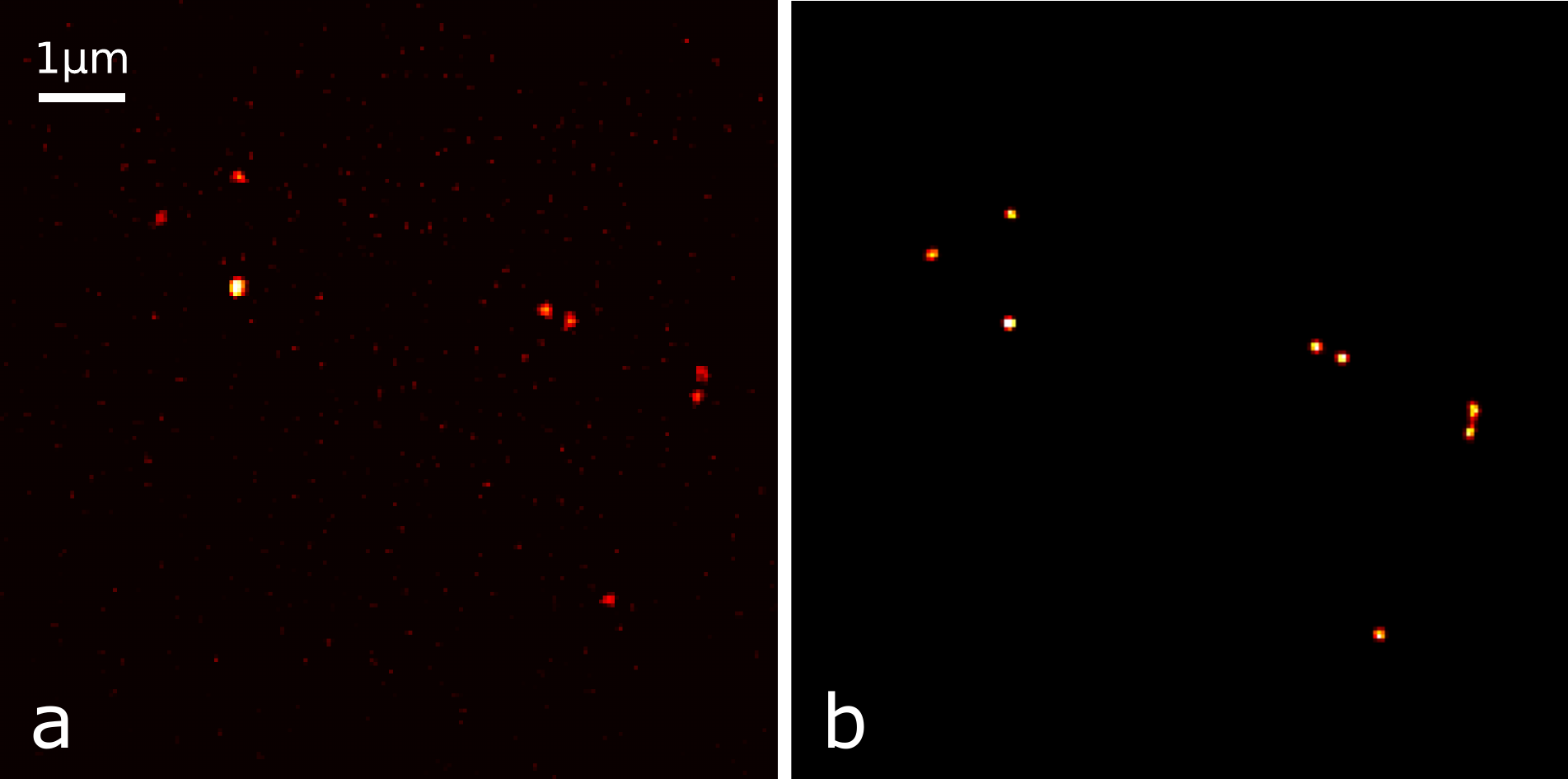}
\caption{\label{fig:WRL_comparison} Comparison of reconstructed images shown in Fig.~\ref{fig:deconvolution} using Wiener deconvolution (a) and Richardson-Lucy deconvolution (b). Richardson-Lucy deconvolution improves the signal to noise ratio.}
\end{center}
\end{figure}

\newpage

\subsection{Blind deconvolution by a phase-retrieval algorithm}

\begin{figure}[!h]
\begin{center}
\includegraphics[width=12cm]{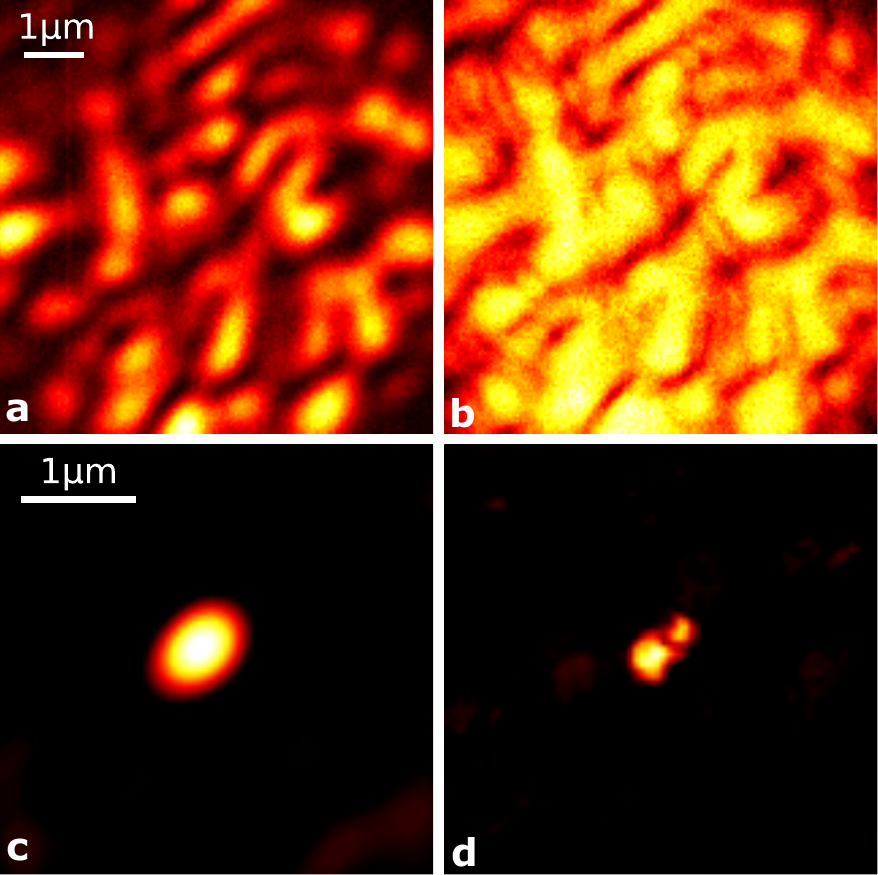}
\caption[]{\label{fig:fienup} Images of the two nearby fluorescent $200~{\rm nm}$ beads obtained with ${\rm NA}=0.33$ and shown in Fig.~\ref{fig:NA_independent}. Speckle image in the linear excitation regime (a) and in the saturated regime (b). 
In the linear regime, the image is reconstructed by Wiener deconvolution (c). Image reconstruction from the saturated speckle image is performed by an iterative phase retrieval algorithm~\citesupp{Fienup_AO_13} (d). The average saturation parameter in b and d is $2.9$ like in Fig.~\ref{fig:NA_independent}. The image in d should be compared to Fig.~\ref{fig:NA_independent}.}
\end{center}
\end{figure}

Blind phase retrieval was performed (Fig.~\ref{fig:fienup}) using a continuous hybrid input output (CHIO) algorithm as described in~\citesupp{Fienup_AO_13}. First, the raw speckle image was periodized using the ``edgetaper'' Matlab function. This function performs a linear interpolation at boundaries of the image and thus allows avoiding artifacts related to fast Fourier transform of data having non-periodic boundary conditions. Next, a difference of Gaussian filter is applied to the auto-correlation: a low-pass Gaussian filter is applied to remove noise from data, and a high-pass Gaussian filter removes the large zero-frequency component due to the power spectrum of the speckle pattern itself. This difference of Gaussian filter with zero mean also allows equilibrating the balance between high spatial frequencies and low spatial frequencies for optimal reconstruction. Finally, $50$ iterations of the CHIO algorithm are run.

\newpage

\subsection{NA-independent resolution}
\begin{figure}[!h]
\begin{center}
\includegraphics[width=12cm]{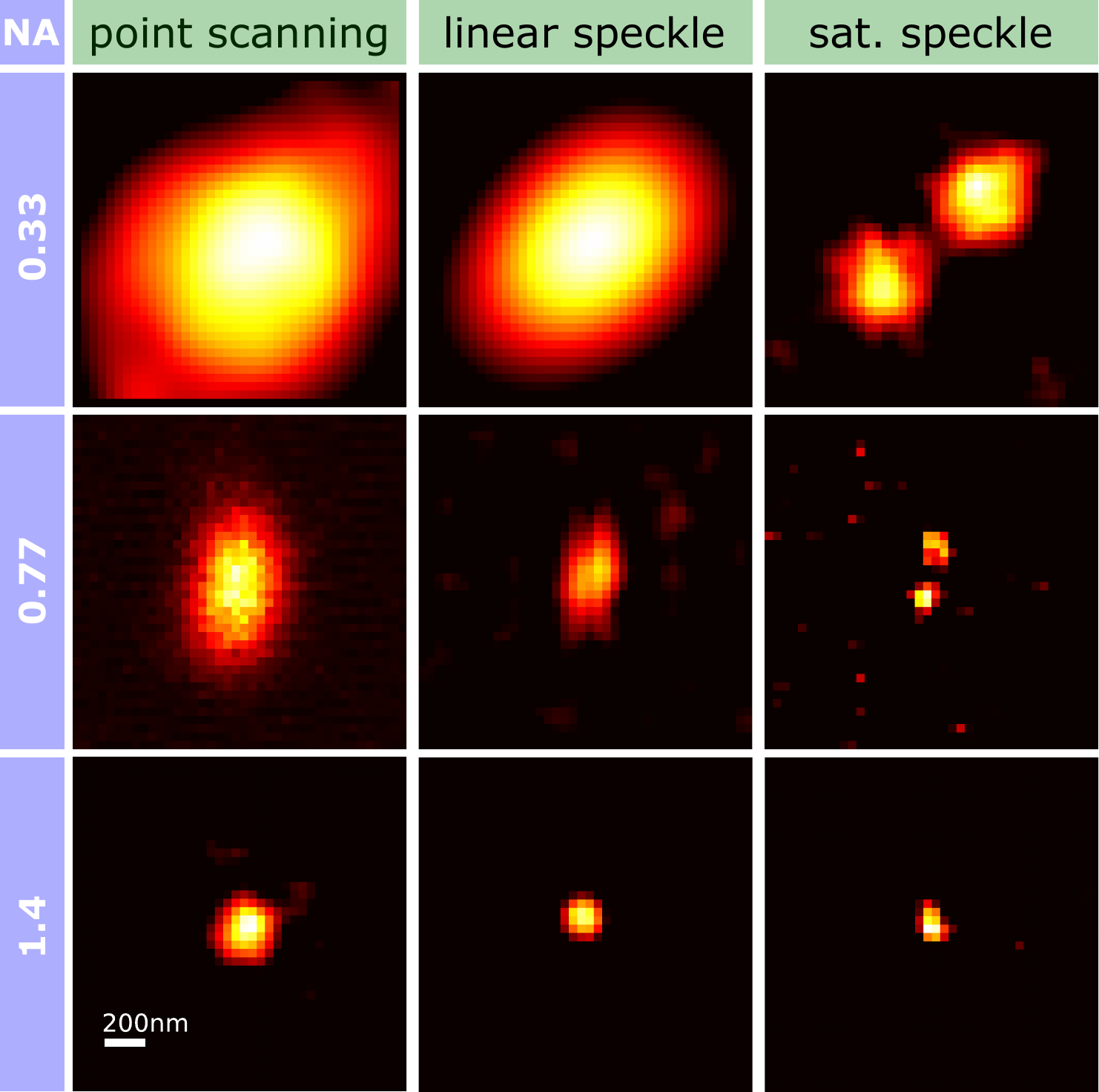}
\caption{\label{fig:NA_independent} Reconstructed super-resolution speckle images (Wiener deconvolution) of fluorescent beads for different NAs. For ${\rm NA}=0.33$, ${\rm NA}=0.77$ and ${\rm NA}=1.4$, the saturation parameters are $\left<s\right>=2.9$, $\left<s\right>=1.4$ and $\left<s\right>=1.6$, respectively. The scale bar is the same for all images.}
\end{center}
\end{figure}

We now discuss and analyze the limit of super-resolution imaging with saturated speckle patterns. 
On the one hand, optical saturation improves resolution thanks to the presence of optical vortices of ${\bf E_\perp}$ (the transverse field).
Circular polarization is chosen in order to minimize the axial field at the center of isotropic optical vortices of same handedness. On the other hand, even in this configuration, saturation degrades the contrast because the intensity at the center of the vortices does not perfectly vanish. The optimal image, featuring the best achievable resolution, should thus be recorded for a saturation level corresponding to the proper balance between resolution improvement -- requiring high saturation levels -- and excitation contrast -- incompatible with too high excitations.

The intensity at the vortex centers follows an exponentially decaying probability density function due to the contribution $I_z$ of the axial field~\citesupp{Pascucci_PRL_16}:
\begin{equation}
\label{eq:pdfIz}
\rho(I_z)=\frac{1}{\overline{I_z}} e^{ -I_z/~\overline{I_z} }
\end{equation}
where $\overline{I_z}$ is the ensemble average of $I_z$ at the center of vortices of ${\bf E_\perp}$.
To achieve super-resolution imaging, the fraction of highly contrasted vortices (remaining dark) in the speckle pattern should be high enough.
As a contrast criterion, we may consider that fluorescence is not significantly excited at the center of a vortex if, there $I_z<I_s$; which limits the saturation level. 
The fraction $f$ of highly contrasted vortices in the population of vortices sharing the same handedness as the polarization is obtained by integrating Eq.~\eqref{eq:pdfIz} over the aforementioned interval:
\begin{equation}
f=\int_{I_z\leq I_s}\rho(I_z) dI_z=1-e^{ -I_s/~\overline{I_z} }
\end{equation}
For $\overline{I_z}=I_s$, the fraction of vortices remaining dark ({\emph i.e.} satisfying $I_z<I_s$) is $\sim 63\%$ among the population of same handedness and $15\%$ among those of opposite handedness (deduced from Eq.~3 in reference~\citesupp{Pascucci_PRL_16}). We may then arbitrarily set $\overline{I_z}=I_s$ as a limit for performing super-resolution imaging.
For a top-hat shaped illumination pupil, $\overline{I_z}=\frac{3}{4}\left(\frac{{\rm NA}}{n}\right)^2 \left<I\right>$ where $n$ is the refractive index of the immersion medium of the objective and $\left<I\right>$ the space average intensity of the speckle pattern~\citesupp{Goodman,Pascucci_PRL_16}. We thus get as a maximum saturation parameter: 
\begin{equation}
\left<s_{max}\right>=\frac{4}{3}\left(\frac{{\rm NA}}{n}\right)^2.
\label{eq:smax}
\end{equation}
Finally, if resolution improves as Eq.~\eqref{eq:rescale} (with $s$ replaced by $\left<s\right>$) we get that the utmost achievable resolution is $\delta x\simeq \frac{\lambda}{2n}\sqrt{\frac{3}{4}}$ (assuming $\left<s_{max}\right>\gg 1$). Interestingly, this limit does not depend on the NA of the imaging lens. In our experimental conditions, Eqs.~\eqref{eq:rescale} and~\eqref{eq:smax} combined together yield $\delta x=152~{\rm nm}$. In practice, we could obviously get super-resolution slightly beyond this limit -- $\delta x\simeq 100~{\rm nm}$ actually limited by the bead size -- suggesting that our theoretical estimate is pessimistic. The reason why Eq.~\eqref{eq:rescale} under-estimates the super-resolving ability of speckles may be because it involves the average saturation factor $\left<s\right>$ while local saturation factors $s$ in a speckle pattern can be much larger. Local high saturation can thus provide super-resolution information with, apparently, high enough signal.

An illustration of the results we obtained using three different NAs is shown in Fig.~\ref{fig:NA_independent}. Resolutions obtained in point-scanning mode as well as after reconstruction from speckle images in the linear and saturated regimes are presented. We observe that both for ${\rm NA}=0.77$ and ${\rm NA}=1.4$, the $100~{\rm nm}$ beads are resolved in the saturated regime $\left<s\right>=1.4$ and $\left<s\right>=1.6$, respectively). For the $0.33~{\rm NA}$, with a saturation parameter of $\left<s\right>=2.9$, it was not possible to reach this same resolution because of the too high pulse energy that would have been required: to get a given resolution $\delta x$ with a given NA, the saturation factor typically scales as: $s\simeq\left(\frac{\lambda}{2\delta x {\rm NA}}\right)^2$. An average saturation parameter larger than $7$ would thus have been required but this was not possible. Using too high energy pulses has, indeed, two drawbacks: - First, it increases photo-bleaching (which increases almost linearly with pulse energy as shown in Fig.~\ref{fig:bleaching}) and thus reduces the statistics of the speckle image - Second, it increases the background signal coming from optics and thus degrades the signal to noise ratio (see Fig.~\ref{fig:saturation}).









\end{document}